\documentclass[american,aps,pra,reprint,superscriptaddress,longbibliography]{revtex4-1}

\usepackage[unicode=true,pdfusetitle, bookmarks=true,bookmarksnumbered=false,bookmarksopen=false, breaklinks=false,pdfborder={0 0 0},backref=false,colorlinks=false] {hyperref}
\hypersetup{colorlinks,linkcolor=myurlcolor,citecolor=myurlcolor,urlcolor=myurlcolor}
\usepackage{braket,colortbl,amsthm,amsmath,amssymb,txfonts,graphicx}
\definecolor{myurlcolor}{rgb}{0,0,0.7}
\usepackage{cleveref}
\usepackage{soul}
\usepackage{subfigure}

\theoremstyle{plain}

\usepackage{ragged2e}  
\usepackage{mathrsfs}
\usepackage{physics}
\usepackage{mwe}
\usepackage{pifont}

\DeclareMathAlphabet{\mathbcal}{OMS}{cmsy}{b}{n}

\usepackage{soul}
\usepackage{pifont}
\usepackage{cancel}

\newtheorem{theorem}{Theorem}

\usepackage{tikz}

\usepackage[normalem]{ulem}

\begin{document}

\title{Exponentially enhanced two-mode multiboson entanglement via phase-modulated tunneling}

\author{Pritam Chattopadhyay}
\affiliation{AMOS and Department of Chemical and Biological Physics,
Weizmann Institute of Science, Rehovot 7610001, Israel}

\author{A. G. Kofman}
\affiliation{AMOS and Department of Chemical and Biological Physics,
Weizmann Institute of Science, Rehovot 7610001, Israel}

\author{Gershon Kurizki}
\affiliation{AMOS and Department of Chemical and Biological Physics,
Weizmann Institute of Science, Rehovot 7610001, Israel}
\date{\today}

\begin{abstract}
The entanglement of quantum systems is commonly restricted by their coupling Hamiltonian and initial state properties. Here, we prove by \textit{exact analysis} of tunnel-coupled bosonic field modes, that \textit{factorized multi-boson} two-mode states can become \textit{fully entangled} via stroboscopic sign flips of the two-mode coupling. Their entanglement can exponentially grow with the number of flips. With the exception of specific prohibitive states that are invariant under such sign-flip control, this effect universally applies to two-mode state preparation. Remarkably, this \textit{linear control} may provide entanglement resources for diverse quantum technological applications by readily available means. 
\end{abstract}

\maketitle

\textbf{Introduction}.--
Since its discovery by Gamow~\cite{Gamow1928}, tunneling has been found to be key to a vast scope of quantum processes~\cite{Gamow1928,belloni2016alpha,ichimaru1993nuclear,jackson1957catalysis,segal2006enhancement,Bardeen1961,Josephson1962,BinnigRohrer1986,SchawlowTownes1958,PhysRevLett.92.200403,PhysRevLett.77.2909,PhysRevA.53.586,PhysRevLett.71.708}. 
Its importance for quantum information processing has been recently recognized by the Nobel Prize~\cite{PhysRevB.35.4682,clarkeprize}. A prominent scenario that underlies diverse applications~\cite{segal2006enhancement,Bardeen1961,Josephson1962,BinnigRohrer1986,SchawlowTownes1958} 
concerns tunneling in a double-well potential. 
The tunneling rate is then hindered not only by the weak coupling through the barrier but also by the \textit{asymmetry/detuning} between the double-well minima.
The degree of tunneling may be very small for large detuning, so that one has to invest a huge amount of energy comparable to the barrier height to activate such processes.

Here we reveal a hitherto unexplored tunneling effect based on \textit{linear dynamical control}: exponential enhancement of the tunneling rate and two-mode entanglement of multiple, non-interacting bosons in asymmetric double-well potentials or their optical analogs in tunnel-coupled waveguides. To this end, we take advantage of tunneling being \textit{a coherent process that can be phase-controlled}. Tunneling was previously shown to result from predominantly \textit{destructive interference of Feynman paths} that cross the potential barrier~\cite{PhysRevLett.77.2909,PhysRevA.53.586}. It should therefore be possible to coherently enhance tunneling by phase modulation of interfering paths on time scales that are much faster than decoherence and dissipation, which tend to render the propagation classical~\cite{PhysRevLett.77.2909,Leggett1984}.

\begin{figure}[htpb]
    \begin{center}
      \subfigure[]{\includegraphics[width=0.48\linewidth]{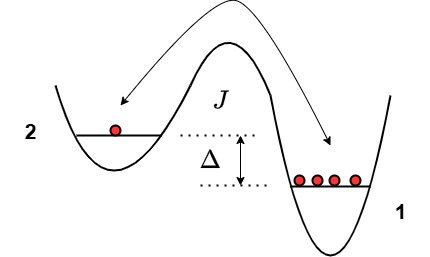}}  
    \subfigure[]{\includegraphics[width=0.48\linewidth]{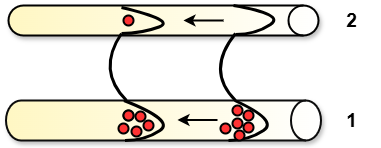}}   
    \end{center}
    \caption{(a) Tunneling in an asymmetric double-well potential with a single level in each well. If well 1 is initially populated by $N$ bosons, and well 2 is empty, then well 2 will be increasingly populated at the expense of well 1. The process is coherent and reversible. (b) Realization of the above scenario for two single-mode ultra-weakly coupled asymmetric waveguides of different widths.} 
    \label{fig1}
\end{figure}

We present an exact solution for controlled multiboson tunneling via a spin-$N/2$ mapping, based upon a theorem we prove for universal entanglement generation under phase-flip control. Undisturbed evolution in this model results in extremely slow tunneling when the two-mode coupling is ultraweak and \textit{no entanglement generation} if the initial single-mode states are factorized (in the number basis) (Fig. \ref{fig1}a). In stark contrast, our exact solution shows that periodic $\pi$-phase jumps (sign-flips) of the coupling at appropriate intervals may exponentially enhance the entanglement and the tunneling rate with the number of phase jumps of initially factorized single-mode states, unless these states possess a symmetry that renders them invariant under the above control.

\begin{figure}[htpb]
 \begin{center}
      \subfigure[]{\includegraphics[width=0.48\linewidth]{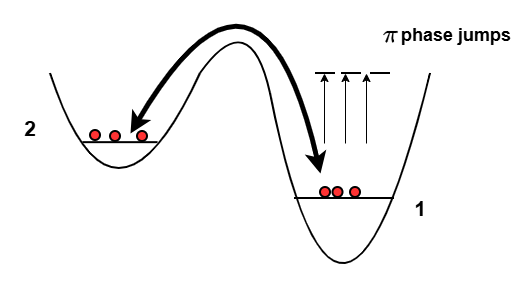}}  
    \subfigure[]{\includegraphics[width=0.48\linewidth]{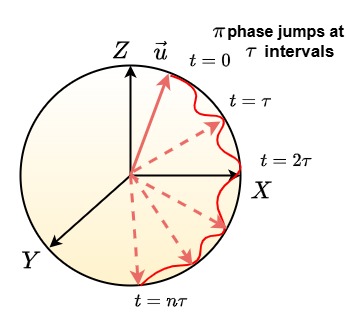}}  
    \subfigure[]{\includegraphics[width=0.55\linewidth]{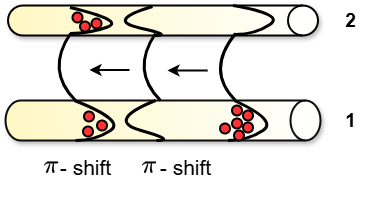}}
    \end{center}
    \caption{(a) $\pi$ phase flips in one well amplify the tunneling of the initially factorized $|N,0\rangle$ state, evolving it to the fully entangled $(|N,0\rangle + |0,N\rangle)/\sqrt{2}$ (NOON) state. (b) Phase jumps on the $N$-dimensional hypersphere.
    (c) Periodically $\pi-$shifted tunnel-coupled waveguides (modified Fig.~\ref{fig1}(b)).}
    \label{fig5aa}
\end{figure}

The most straightforward realization of this \textit{entirely linear entanglement control} is envisaged in weakly coupled photonic waveguides (Fig. \ref{fig1}b) whose coupling is periodically phase-modulated along the waveguide (Fig. \ref{fig5aa}c). We foresee advantageous applications of this scheme as an entanglement resource of quantum information processing~\cite{nielsen2010quantum,Kimble2008,PhysRevA.65.032323,PhysRevLett.78.2275,sur2025molecular,PhysRevLett.80.5239} and diverse quantum technologies~\cite{Smerzi1997,Albiez2005,Dalfovo1999,leonhardt1997measuring,Boto2000,Dowling2008,Schwinger1952,besse2020realizing,PhysRevA.61.062311,PhysRevLett.78.3221,kuang2025perfect,RevModPhys.79.135,PhysRevLett.130.263602,RevModPhys.89.015005,lee2025photonic,PhysRevA.78.063811,PhysRevA.71.043805} (see Discussion).

\textbf{Undisturbed tunneling between two bosonic modes}.--

The two bosonic modes in our scenario may be viewed as the energy-mismatched (mutually-detuned) ground levels of two asymmetric potential wells that are coupled by tunneling through the potential barrier between the wells. We assume that the boson \textit{inter-particle interaction} is \textit{negligibly small compared to the tunneling strength}, so that the \textit{dynamics is linear}, unlike that of commonly investigated Kerr-nonlinear Bose condensates~\cite{Smerzi1997,Albiez2005,Dalfovo1999}. Higher levels in these wells, associated with additional bosonic modes, are assumed to be much further detuned and are therefore excluded from our analysis. 

Then the Hamiltonian is given by~\cite{Smerzi1997,PhysRevA.20.539,Shore1990} 
\begin{equation} \label{a1}
    \hat H = J \hat L_x + \Delta \hat L_z, 
\end{equation}
where we assume for concreteness that $J > 0$ and $\Delta \geq 0$. Here, $J$ is the resonant tunneling strength~\cite{agarwal2012quantum}, $\Delta$ is the detuning (energy mismatch) between the modes, and $\hat L_i$ ($i$ = $x, y, z$) are the collective N-boson angular momentum operators. In the Jordan-Schwinger representation, they are expressed in terms of the annihilation and creation bosonic operators, $\hat{a}_{1}$, $\hat{a}_{1}^\dagger$  and $\hat{a}_{2}$, $\hat{a}_{2}^\dagger$ of modes 1, 2 \cite{Schwinger1952},
\begin{align} \nonumber \label{a2}
& \hat{L}_{x}=\frac{\hat{a}_{1}^{\dagger} \hat{a}_{2}+\hat{a}_{2}^{\dagger} \hat{a}_{1}}{2}, \quad \hat{L}_{y}=\frac{\hat{a}_{1}^{\dagger} \hat{a}_{2}-\hat{a}_{2}^{\dagger} \hat{a}_{1}}{2 i}, 
 \hat{L}_{z}=\frac{\hat{a}_{1}^{\dagger} \hat{a}_{1}-\hat{a}_{2}^{\dagger} \hat{a}_{2}}{2} . 
\end{align}

As a simple case, consider N bosons shared by the two modes. There are $N + 1$ collective states $|n\rangle$ ($n = 0, 1, . . . , N$), which correspond to n (respectively, $N - n$) bosons in mode 1 (2) that occupy the ground levels of the two wells. These two-mode states $|n, N-n \rangle \equiv |m, l\rangle$ are eigenstates of $\hat L_z$  
\begin{equation} \label{a3}
|n, N-n \rangle \equiv |m, l\rangle; \quad \hat{L}_z |m, l\rangle = m |m, l\rangle; \quad  m=n-l, \quad l =N/2
\end{equation}
$l$ being the value of the fictitious $N$-boson angular momentum. The above description holds regardless of the shape and the width of each potential well, as long as the higher levels of the double-well potential remain unpopulated. 

The evolution of the initial N-boson wave function under Hamiltonian (1), 
\begin{equation} \label{a4}
|\psi(t)\rangle= e^{-iHt} |N,0 \rangle = \sum_{m=-l}^{l} c_{l,m}(t) |m, l\rangle,
\end{equation}
is isomorphic to the rotation caused by the unitary evolution operator 
\begin{equation} \label{a5}
\hat U(t) = e^{-i \Omega t (\hat{L}_x \sin \theta + \hat{L}_z \cos \theta)}, \quad \Omega=\sqrt{J^{2}+\Delta^{2}}, \theta =\arctan (J/\Delta).
\end{equation}

The initial product state $|N,0\rangle \equiv | m=l, l \rangle$, wherein all bosons are initially in mode 1, evolves via \eqref{a5} to the two-mode \textit{entangled form} \eqref{a4} of superposed $m-$states with fixed $l=N/2$ having the probability amplitudes~\cite{varshalovich1988quantum,edmonds1996angular} 
\begin{equation}\label{a9}
c_{l,m}(t)=\langle l,m|\hat U(t)|l,l\rangle =\sum_{m^\prime} d_{mm^\prime}^{(l)}(\theta) e^{-im^\prime \Omega t} 
d_{lm^\prime}^{(l)*}(\theta),
\end{equation}
where $ d_{mm^\prime}^{(l)}(\theta)$ are the real-valued Wigner d-matrix elements dependent on the rotation angle $\theta$ (see SI I).

The survival probability of the initial $|N,0\rangle$ state 
$P_N (t)$ which is complementary to the tunneling probability, $P_{tun}^{(N)} = 1- P_N (t)$, can be evaluated to be (see SI I)
\begin{equation}\label{a13}
P_{N}(t) \approx \left(1-\frac{J^{2}}{\Omega^{2}} \sin ^{2} \frac{\Omega t}{2}\right)^{N}. 
\end{equation}

Off-resonance, i.e. for $\Delta \neq 0$, complete transfer in the $|n, N-n\rangle$ basis is impossible. However, according to Eq.~\eqref{a13}, the population of the initial state can decrease from 1 to a very small value near resonance: for $\Delta^{2} \ll N J^{2}$ it approximately exhibits periodic Gaussian revivals of the survival probability:  
\begin{equation} \label{a16}
P_{N}(t) \approx e^{-N J^{2}\left(t-t_{2 k}\right)^{2} / 4}, \quad\left|t-t_{2 k}\right| \ll \Omega^{-1}, \quad (k=1,2, \ldots),
\end{equation}
where $t_k =k\tau$. This expression is independent of $\Delta$, i.e., the effects of the detuning are insignificant in this limit. It is close to the resonant case wherein the population is $P_{N}(t)=\cos ^{2 N} \frac{\Omega t}{2}$, so that $P_N (t)$ vanishes at times when all the bosons are transferred to mode 2. In the opposite limit of large detuning, $\Delta^{2} \gg N J^{2}$, the decay of the survival probability is always small.

Equations \eqref{a13}-\eqref{a16} show that the effective strength of the coupling of the state $|N,0 \rangle$ to the other states is \textit{cooperatively enhanced} to the value $\sqrt{N} J$. This hitherto unknown \textit{collective transfer enhancement} implies that the transfer rate can be amplified by a factor of $N$ compared to a single-boson rate. This process is coherently reversible, unlike the well-known open-system anti-Zeno effect~\cite{kurizki2022thermodynamics,PhysRevLett.87.270405,kofman2000acceleration,biagioni2008experimental}.  
Since the total number operator $\hat N = a_1^\dagger a_1 +a_2^\dagger a_2$ commutes with the Hamiltonian \eqref{a1}, an arbitrary initial superposition of $|N,0\rangle$, i.e. $N$-boson numbers in one mode and $0$ in the other having amplitudes $A_N$ evolves as
\begin{equation}\label{psit}
    |\psi(t)\rangle = e^{-i\hat H t} \sum_N A_N |N, 0\rangle = \sum_{l \in \{0, \frac{1}{2}, 1, \dots\} } \Tilde{A}_{2l}
    \sum_{m=-l}^{l (N)} c_{l,m}(t) |l,m\rangle.
\end{equation}
where $\Tilde{A}_{2l}$ originate from the initial $A_N$ distribution whereas $c_{l,m}$ are the same as in \eqref{a9}. 
Equation \eqref{psit} describes the independent evolution of each N-sector via tunneling which is a \textit{linear passive operation}.

\textbf{Tunneling control by periodic phase jumps}.--
For a fixed $N$ and large $\Delta$, $P_{tun} = 1-P_N(t)$ is bounded by an exponentially small value, as can be easily seen from \eqref{a13}. 
Let us show that, irrespective of the detuning $\Delta$, \textit{the tunneling expressed by \eqref{psit} can become complete} under appropriate phase jumps. Since the undisturbed multiboson tunneling process for a fixed initial $N$ is a \textit{highly restricted rotation} around an axis $\vec{u} = \sin \theta \hat x + \cos \theta \hat z$ of the $N-$ dimensional hypersphere,  we aim at lifting the restriction for $\Delta >> J$. 

To this end, the mode detuning is modulated in time, so that the Hamiltonian (1) changes to
\begin{equation}\label{b1}
\hat{H}=J \hat{L}_{x}+[\Delta+\delta(t)] \hat{L}_{z}.
\end{equation}
where $\delta(t)$ is the intermode detuning change due to a control field. In the frame rotating around the $z$ axis at the frequency $\delta(t)$, we can write
\begin{equation}\label{b2}
|\psi(t)\rangle=e^{-i \phi(t) \hat{L}_{z}}|\chi(t)\rangle, \quad \phi(t)=\int_{0}^{t} d t^{\prime} \delta\left(t^{\prime}\right),
\end{equation}
Upon transforming to the frame that rotates by $\phi(t)$ around $\hat L_z$;
the Hamiltonian acting on $|\chi(t)\rangle$ becomes
\begin{equation}
    \tilde{H}(t) = J[\cos\phi(t) \hat L_x + \sin\phi(t) \hat L_y] + \Delta \hat L_z.
\end{equation}

Let us take $\delta(t)$ to consist of identical short (\textit{stroboscopic}) pulses  that produce $\pi$-phase shifts (flips) at intervals $\tau$ (Fig.~\ref{fig5aa}a,b), so that, on neglecting the width of the pulses, $\phi(t)=\pi[t / \tau]$,
where $[x]$ denotes the integer part of $x$.  At even- or odd numbered intervals $t_{2 k} \equiv 2 k \pi / \Omega, t_{2 k + 1} \equiv (2 k +1) \pi / \Omega,$ the propagator $\hat{U}\left(t_{2 k}\right)$, rotates the system, respectively, in the $x-z$ plane, rendering (SI II)
\begin{subequations}\label{b21gen}
\begin{equation}\label{b21}
\hat{U}\left(t_{2 k}\right)=e^{i 4 k \theta \hat{L}_{y}},
\end{equation}
\begin{equation} \label{b22}
\hat{U}\left(t_{2 k+1}\right)=e^{-i \pi \hat{\tilde{L}} \cdot \vec{u}_{k}}, \quad \vec{u} =(\sin (2 k+1) \theta, 0, \cos (2 k+1) \theta),
\end{equation}
\end{subequations}
where $\theta$ is defined in \eqref{a5}. Since we have chosen $\cos\phi(t)=(-1)^{\lfloor t/\tau\rfloor}$ and $\sin\phi(t)=0$, 
the Hamiltonian alternates in sign:
\begin{equation}\label{hstrosb}
    \tilde{H}(t) = J(-1)^{\lfloor t/\tau\rfloor} \hat L_x + \Delta \hat L_z.
\end{equation}

Here we assume identical switching intervals and exact $\pi$-phase flips. Errors can reduce the constructive interference and hence the attainable tunneling and entanglement enhancement. 
Details of the robustness to small deviations in the switching and the phase parameter are deferred to SI V.

Physically, this corresponds to \textit{alternating tunneling direction} every $\tau$. For the $|N,0\rangle$ component the survival probability is then 
\begin{equation}\label{b29}
P_{N}\left(t_{k}\right)= \cos ^{2 N}(k \theta) \quad(k=0,1, \ldots).
\end{equation}

Thus, control by $\pi$ phase shifts (sign flips) does not significantly affect the rate of decay of $P_{N}(t)$ for small detuning $\Delta^{2} \ll N J^{2}$. By contrast, in the large-detuning limit, $\Delta^{2} \gg N J^{2}$, where the unperturbed tunneling is insignificant, this phase-flip control can result in \textit{almost complete decay} of the initial state. This hitherto unexplored solution is here deemed the closed-system \textit{collective anti-Zeno effect} (CAZE), since it enhances the decay rate from zero to a finite value that scales with $N$. In particular, for large detuning $\Delta \gg J$, Eq. \eqref{b29} yields approximately,
\begin{equation}\label{b30}
P_{N}\left(t_{k}\right) \approx \exp \left(-k^{2} \frac{N J^{2}}{\Delta^{2}}\right), \quad k \ll \Delta / J, 
\end{equation}
which exhibits \textit{collective $k^2N$ scaling of the decay exponent}. It should be contrasted with the open-system anti-Zeno effect of accelerated decay via coupling to a thermal bath or continuum~\cite{kurizki2022thermodynamics,PhysRevLett.87.270405,kofman2000acceleration}.

\begin{figure*}[htpb]
 \begin{center} 
 \subfigure[]{\includegraphics[width=0.46\linewidth]{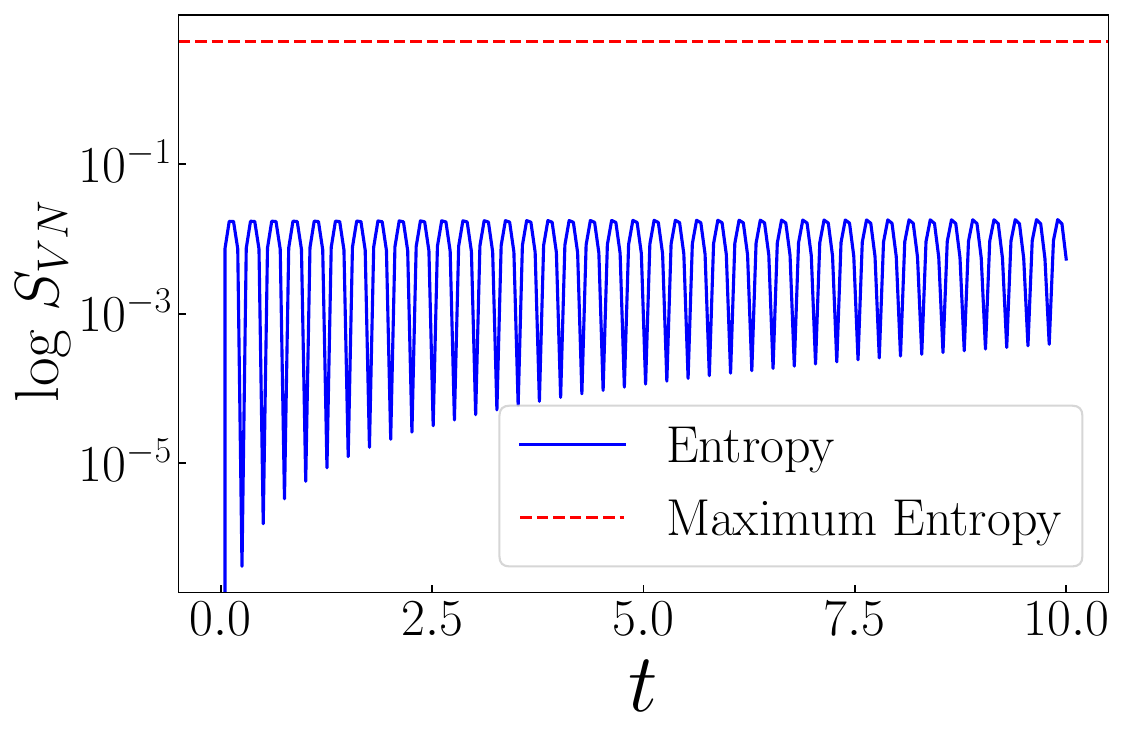}}
    \subfigure[]{\includegraphics[width=0.46\linewidth]{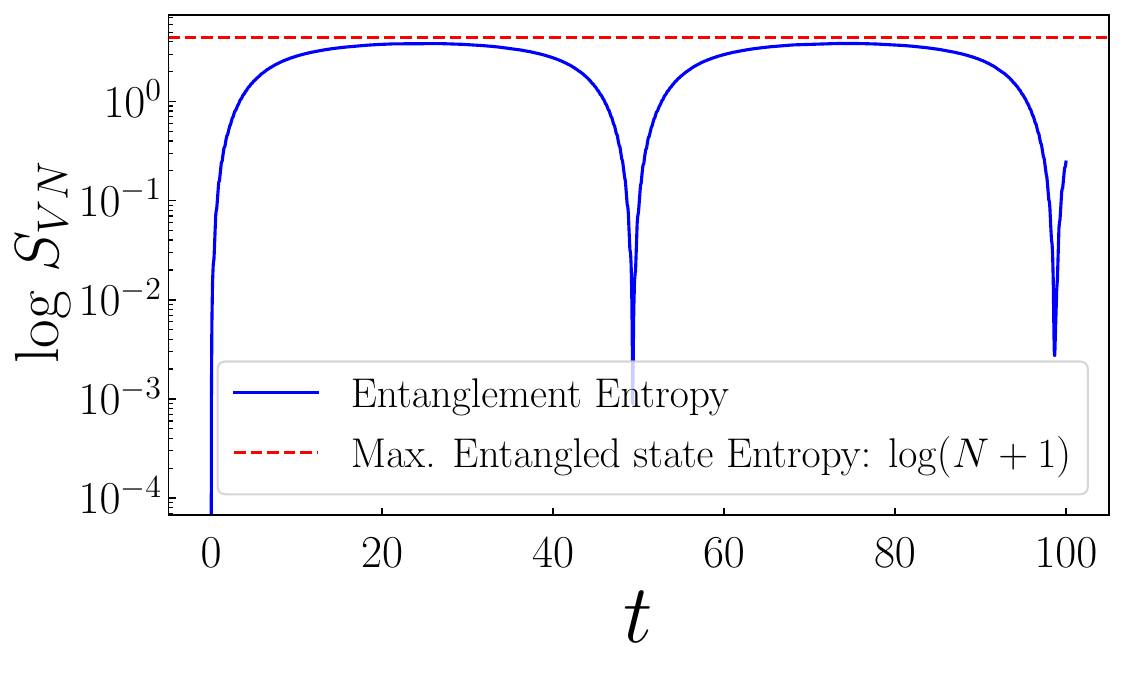}}\\
    \subfigure[]{\includegraphics[width=0.46\linewidth]{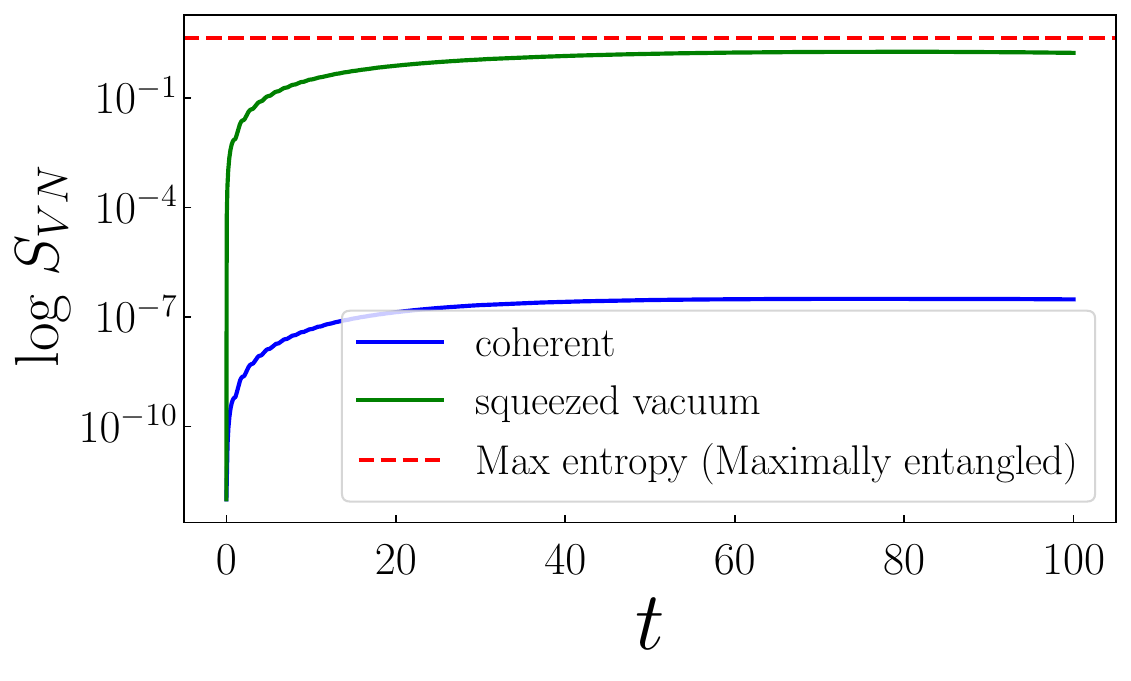}}
    \subfigure[]{\includegraphics[width=0.46\linewidth]{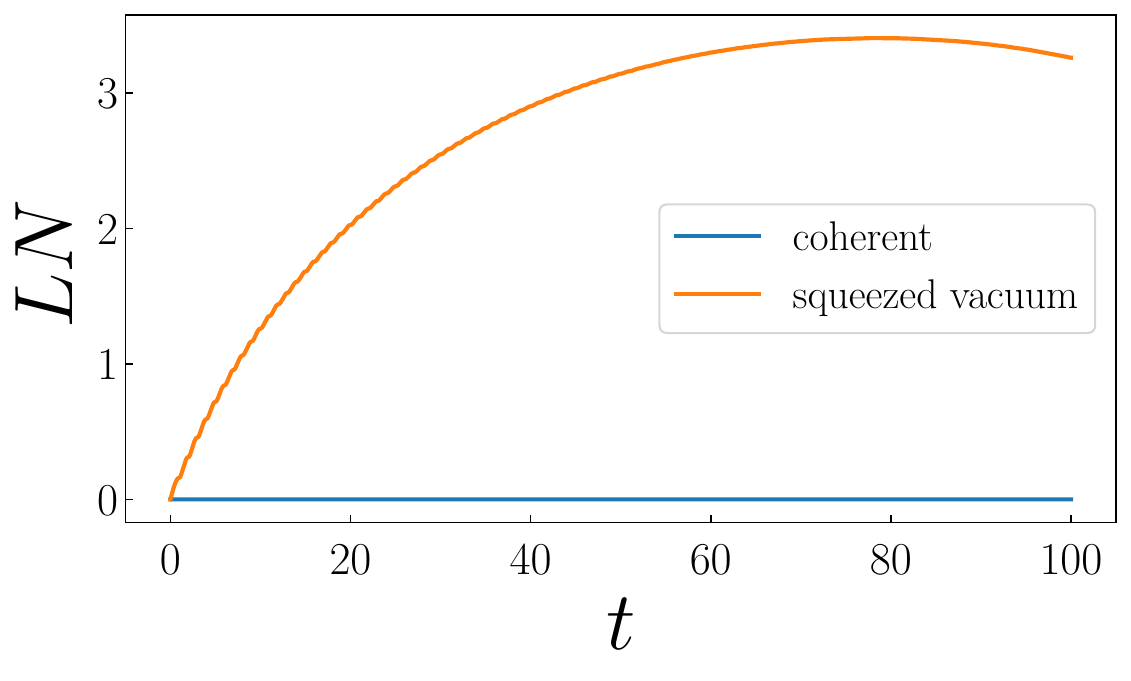}}
    \end{center}
\caption{(a) Von-Neumann entanglement entropy ($S_{VN}$) evolution for undisturbed tunneling with $J=1$, $\Delta =100$, $N=20$, in time units of $\pi/\sqrt{\Delta^2+ J^2}$, is restricted to values well below full entanglement. (b) Entanglement entropy as a function of time under phase flips. For (a), (b) the initial state is $\vert N,0\rangle$. (c)  Comparison between the evolution of entanglement entropy under phase-flips of coherent state with amplitude $\alpha=2$ and squeezed vacuum states (squeezing parameter $r=2$) with $J=1$, $\Delta =100$. (d) Same as (c) for logarithmic negativity $LN$ as a function of time under phase flips. Both (c) and (d) show that entanglement is severely restricted for spin-coherent states but exponentially enhanced for squeezed states.}
    \label{fig5}
\end{figure*}

\textbf{Generation and enhancement of entanglement}.--
Eqs.\eqref{a4}-\eqref{a9} describe a state that becomes entangled in the $|n, N-n\rangle$ basis via tunneling, having started from the factorized state where only mode 1 is occupied. 
The general question we ask is: \emph{Under what conditions does the stroboscopic control \eqref{b21gen}-\eqref{hstrosb} of an arbitrary pure two-mode state generate or enhance two-mode entanglement within a finite number of stroboscopic steps?}
The answer is given by the following theorem.
\begin{theorem} 
Any initial two-mode state that exhibits coherent off-resonant tunneling dynamics under the unperturbed evolution undergoes, under phase-jump control, stroboscopic SU(2) dynamics equivalent to resonant coherent oscillations/tunneling. In particular, for the initial Fock state $|N,0\rangle$, the evolution yields $P_N(t_k)=\cos^{2N}(k\theta)$, mimicking resonant oscillations.
Moreover, a necessary condition for the generation or enhancement of two-mode entanglement is that the initial state is nonclassical and that the stroboscopic evolution populates both modes.
 
\end{theorem}

The proof is given at End Matter.

To \textit{quantify the two-mode entanglement}, we resort to the von Neumann entropy of the reduced density matrix for mode $1$ or mode $2$. 
This entropy grows due to tunneling-induced population mixing. An elaborate calculation (SI III) yields the entanglement entropy bound of the N-sector state 
\begin{equation}
    S_N(t) = -\sum_m P_N(m,t)\ln P_N(m,t)
    \;\approx\;
    \frac{1}{2}\ln(2\pi \,\sigma_{\mathrm{N}}^2)
    + \ln \eta_N,
    \label{eq:entropy_bound}
\end{equation}
where the effective variance $\sigma_{\mathrm{N}}^2$ is fixed for unperturbed evolution (Fig.~\ref{fig5}a) but scales with $k$ under sign-flip modulation, and $\eta_N \sim O(1)$ accounts for finite support corrections (SI II).

Let us assume, for example, that the initial state is $|r,0\rangle$: squeezed vacuum with squeezing parameter $r$ in mode 1 and the vacuum state in mode 2.
Under stroboscopic sign-flip modulation control, there is a drastic monotonic increase in the entropy (Fig.~\ref{fig5}b). Subtracting the unperturbed entropy bound from its perturbed (modulated) counterpart yields the entropy enhancement due to modulation after $k$ cycles (SI III)
\begin{eqnarray}\label{entropychange} \nonumber
    \Delta S_k 
   & = & S_{\mathrm{mod}}(t_k) - S_{\mathrm{unmod}}(t), \\
    & \gtrsim &
    \frac{1}{2}\ln k 
    + 
    \frac{1}{2}
    \ln\!\left(
        \frac{
            \sin^2\theta
        }{
            \sin^2\theta + \cos^2\theta(\sinh^2 r + 1)
        }
    \right),
\end{eqnarray}
where $\theta$ is as in \eqref{a5}.

The following asymptotic limits of \eqref{entropychange} transpire: a) For weak squeezing ($r\ll1$); $\sinh^2 r\simeq r^2$, $\Delta S_k\simeq \frac{1}{2}\ln k$, recovering the entropy scaling for initial $|N,0\rangle$ state. b) For strong squeezing ($r\gg1$); $\sinh^2 r\gg1$, the prefactor in the last logarithmic term of \eqref{entropychange} decreases, indicating that the unperturbed (unmodulated) entropy is already high, while the modulation-induced growth retains the $\tfrac{1}{2}\ln k$ scaling, albeit with smaller relative enhancement. This result ensures reaching full two-mode entanglement after a finite ($k$) number of steps, regardless of how small the initial-state squeezing $r$ was.

Indeed, at long times, as $k \rightarrow \infty$, the phase-modulated entanglement entropy approaches:
\begin{equation}\label{21eqna}
    S_{\text{mod}}(t) \to \ln(2l + 1)
\end{equation}
which is the maximal possible mode-entanglement entropy for a two-mode bosonic system with total boson number \( N = 2l \), associated with the fully-entangled state~\cite{Dowling2008}. The maximal entropy in Eq.~\eqref{21eqna} is reached when the reduced density matrix becomes uniform, $P_N(t)=1/(N+1)$, corresponding to equal population of all $\vert n, N-n \rangle$ states. This requires initial states with sufficient support over the controlled eigenmodes and, in particular, exclusion of the symmetry-protected invariant (spin-coherent) states, which do not spread under the stroboscopic control dynamics.

An alternative measure of entanglement is logarithmic negativity~\cite{RevModPhys.81.865,1y1pzqhh,PhysRevLett.95.090503,adesso2007entanglement,PhysRevA.65.032314,PhysRevA.65.032323} (Fig~\ref{fig5}(d)) that depends only on $P_N(n,t)$
\begin{equation}\label{lognegativity}
    LN(t)
=2\log_2\!\left(\sum_{n=0}^{N}\sqrt{P_N(n,t)}\right).
\end{equation}

The regime $\Delta \gg J$ highlights the strongest impact of the protocol: whereas uncontrolled tunneling and entanglement are strongly suppressed, stroboscopic $\pi$-flips restore substantial population transfer and entanglement growth. As $\Delta$ approaches $J$, this advantage diminishes because the uncontrolled tunneling and entanglement are then already significant and largely insensitive to $\Delta$ even without control. Thus, the protocol is most effective for large-detuning.

\textbf{Discussion}.--
We have obtained an exact solution for tunnel-coupled mode-sharing multiple-boson dynamics. This solution has unraveled a hitherto unknown effect of giant entanglement amplification and tunneling speedup via $\pi-$phase flips of the state vector at the rate that corresponds to the frequency mismatch/detuning $\Omega$ of the two modes. The analysis resorts to the mapping of the problem onto that of the $N-$dimensional angular-momentum vector that precesses on the hypersphere under the action of a fictitious magnetic field. This precession can be dramatically amplified by  $k$ sign flips, a case that can be \textit{solved exactly}, although the Hamiltonian consists of two non-commuting terms. This comes about since the precession axis of the state vector jumps around with each phase flip and eventually fills the entire surface of the $N-$dimensional hypersphere.  The problem bears analogy to stepwise coherent excitation of a system with $N$ equidistant levels~\cite{Shore1990}.  

In general, the entanglement of two-mode states via unitary evolution is restricted by the strength and nonlinearity of their coupling~\cite{Sorensen2001} or, alternatively, by \textit{non-unitary} measurement-induced effects~\cite{Chou2005,sherson2006quantum}. Here, remarkably, we resort to linear control via periodic phase flips to exponentially boost entanglement of the initial factorized state. Our exact SU(2)-based treatment and entanglement theorem are restricted to the two-mode case, leaving genuine multipartite extensions as an interesting direction for future work.

Importantly, the results of our analysis differ drastically from those of the classical analog: two coupled waveguides of different widths redistribute the intensity ratio as $I\sim d^2$, $d$ being the spatial modulation period~\cite{Snyder1983,Yariv1973}. Thus, our predicted entanglement generation and exponential enhancement with $k$ sign flips is a distinctly quantum effect, without classical counterpart. 

Our description goes well beyond the domain of applicability of the universal Kofman-Kurizki (KK) formula that describes, among many processes~\cite{kurizki2022thermodynamics}, the perturbative dynamical control of bound-state tunneling to a continuum of states~\cite{PhysRevLett.87.270405}. Depending on the spectrum of the control process, the KK formula yields either decay/tunneling slowdown (alias the quantum Zeno effect – QZE) or, conversely, its speedup (dubbed the anti-Zeno effect – AZE)~\cite{kofman2000acceleration}. This formula has been shown to describe macroscopic tunneling control in SQUIDS~\cite{PhysRevLett.92.200403} and in bosonic junctions~\cite{PhysRevLett.100.220403}: it treats tunneling from a bound state to the continuum as a \textit{nearly-irreversible process}. By contrast, we here study tunneling between two modes of a \textit{closed system}, which is a \textit{reversible process}. Nevertheless, the exact solution for the dynamical control of tunneling has been shown here to yield giant analogs of AZE. These analogs have been found to be amplified by the many-body collective (bosonic) character of tunneling, which is a hitherto unknown feature.

The detuning in our model is simply the mismatch between the two mode frequencies/energies. In the two realizations discussed in the work, this corresponds, respectively, to the asymmetry of the two wells or to the propagation-constant mismatch of two asymmetric waveguides (Fig.~\ref{fig1}b). We assume that the modes are detuned because of well asymmetry or different waveguide widths, while the intermode coupling $J$ is independently set by the barrier transparency or, in the photonic implementation, by the overlap of the evanescent tails. This makes the condition $\Delta \gtrsim J$, and in particular $\Delta \gg J$, physically natural. In the photonic-waveguide realization, moreover, the control does not require dynamically reaching resonance. One only needs weakly coupled asymmetric waveguides subject to periodically spaced $\pi$-phase shifts, separated by intervals $\pi c/\Omega$ (equivalently $\pi/\Omega$ in time units). The required phase-coherent control and stroboscopic $\pi$-phase updates ($\tau \sim \pi/\Omega$) are within reach of existing Josephson-junction and photonic platforms~\cite{PhysRevLett.95.010402,n2y32bmz,taghavi2025ghz}.

This scenario may give further impetus to the emerging trend of exploring and implementing quantum electrodynamic effects in waveguides~\cite{Ciccarello2024,PRXQuantum.6.010342}. It is particularly promising in the context of quantum metrology~\cite{Dowling2008}, lithography~\cite{PhysRevLett.109.103602}, and quantum communication by means of entangled photonic NOON states, which can serve as \textit{deterministic quantum repeaters}, unlike the usual probabilistic quantum repeaters~\cite{Ladd2006}. Another intriguing potential application of the entangled NOON states obtained by the present scheme is for supersensitive microscopy~\cite{PhysRevLett.109.103602,PhysRevLett.112.103604,PhysRevA.110.013715}.

The predicted effects bode well with novel designs of integrated solid-state and photonic structures, comprised of either asymmetric bosonic Josephson junctions or weakly coupled photonic waveguides with periodic phase modulation. In integrated photonics, programmable phase control is already standard. Reconfigurable quantum photonic circuits based on phase shifters and directional couplers have demonstrated programmable quantum interference with high visibility, showing that phase control can be exerted while preserving quantum coherence~\cite{n2y32bmz}. GHz-frequency phase shifters have now been demonstrated to be faster than relevant coherence windows of on-chip photonic circuits~\cite{taghavi2025ghz}.
Leveraging such advancements, the predicted effects can open a new route towards quantum multiphoton state control.

Realistically, some dissipation is always present in the system. In the simple non-Hermitian approach, the energy eigenvalues are then modified by $i \Gamma$, where $\Gamma$ is the dissipation rate. This should lead to damping of the coherent oscillations in Eq.~\eqref{b29}. In low-leakage optical waveguides, we may attain $\Gamma \lesssim$ MHz~\cite{taghavi2025ghz} that would lead to weak damping in Eq.~\eqref{b29} $\sim e^{-\Gamma t}$ for oscillations on a GHz scale of currently available phase-shifters~\cite{n2y32bmz} against stochastic errors.

\textbf{Acknowledgement}.--  P.C. acknowledges the support from the International Postdoctoral Fellowship from the Ben May Center for Theory and Computation. G.K. acknowledges the support from a research grant from Magnus Konow in honour of his mother Olga Konow Rappaport.


%

\onecolumngrid
\newpage
\vspace{10ex}
\begin{center}
   \Large \textbf{End Matter}
\end{center}

\section*{ Proof of Theorem 1}

Let the initial state be an arbitrary two-mode pure state $|\psi(0)\rangle = \sum_{N=0}^\infty |\psi_N(0)\rangle$, where each $|\psi_N(0)\rangle$ belongs to the $N$-boson sector $\mathcal{H}_N = \mathrm{span}\{|n,N-n\rangle\}_{n=0}^N$. Since the Hamiltonian conserves the total boson number $\hat N$, the evolution decomposes into independent dynamics in each $N$-sector
\begin{subequations}
\begin{equation}
|\psi(k\tau)\rangle = \sum_N \hat U_N(k\tau)\,|\psi_N(0)\rangle.
\end{equation}
We now make explicit the structure of the evolved state in each $N$-sector. 
Expanding in the Dicke basis $\{|l,m\rangle\}_{m=-l}^{l}$ with $l=N/2$, one has
\begin{equation}
\hat U_N(k\tau)|\psi_N(0)\rangle 
= \sum_{m=-l}^{l} c_{l,m}(t)\,|l,m\rangle,
\end{equation}
\end{subequations}
where the amplitudes $c_{l,m}(t)$ are determined by the SU(2) evolution.
Upon tracing out mode 2, the reduced density matrix of mode 1 in the $N$-sector becomes diagonal in the number basis,
\begin{equation}\label{thm1}
 \rho_1^{(N)}(t=k\tau) = \sum_{n=0}^N P_N(n,t) \, |n\rangle\langle n|,  
 \quad    
 P_N(n,t)\equiv |c_{l,m}(t)|^2,
\end{equation}
where $l=N/2$ and $m=n-l$. The reduced state $\rho_1^{(N)}(t)$ is pure only if a single amplitude $c_{l,m}(t)$ is nonzero; otherwise it is mixed. Since the global state remains pure under unitary evolution, a mixed reduced state implies two-mode entanglement. Therefore, entanglement is generated or enhanced if, in a sector $N_0\equiv 2l_0$, the evolved state $\hat U_{N_0}(k\tau)|\psi_{N_0}(0)\rangle$ acquires at least two nonzero amplitudes with distinct $m$.

The unitary evolution \eqref{b21}, \eqref{b22} corresponds to SU(2) rotations in each sector:
\begin{equation}\label{thm2}
    \hat U_N(k\tau) = \hat R_k \cdots \hat R_2 \hat R_1, 
    \quad 
    \hat R_p = \exp\!\bigl(-i \theta_p \,\hat L_{u_p}\bigr).
\end{equation}
The subgroup $G\subset \mathrm{SU}(2)$ is generated by rotations about two non-collinear axes (due to alternating $\hat L_x$ sign and $\hat L_z$ evolution). Because the spin-$l$ representation of $\mathrm{SU}(2)$ is irreducible, it admits no nontrivial invariant subspaces. However, irreducibility alone does not guarantee that a given initial state spreads over multiple $|l,m\rangle$ components under a finite sequence of rotations.
To establish spreading, we note that the stroboscopic evolution is generated by rotations about at least two non-collinear axes. Therefore, the unitary operators $\hat R_p$ do not commute and generate a non-Abelian subgroup of $\mathrm{SU}(2)$. 
For any initial state $|\psi_N(0)\rangle$ that is not an eigenstate of all generators $\hat L_{u_p}$, there exists at least one rotation $\hat R_p$ such that $\hat R_p |\psi_N(0)\rangle \not\propto |\psi_N(0)\rangle$.
Hence, the state acquires components along multiple $|l,m\rangle$ basis states.
Iterating this procedure, the successive action of non-commuting rotations produces a superposition involving at least two distinct $m$ values after a finite number of steps, unless the initial state is invariant under the generated subgroup (e.g., symmetry-protected spin-coherent states aligned with the rotation axes).
Therefore, for all non-invariant initial states, there exists a finite $k$ such that the evolved state $\hat U_N(k\tau)|\psi_N(0)\rangle$ has at least two nonzero amplitudes $c_{l,m}(t)$.

In particular, for the initial Fock state $|N,0\rangle$, which corresponds to the highest-weight state $|l,l\rangle$, the rotations $\hat R_p$ are not aligned with $\hat L_z$. Therefore, the state is not invariant and undergoes nontrivial redistribution, yielding $ P_N(t_k)=\cos^{2N}(k\theta)$, which reproduces the functional form of resonant coherent oscillations with accumulated angle $k\theta$.

Explicitly, the amplitudes evolve as
\begin{equation}
\mathbf{c}(k\tau) 
= D^{(l)}(\hat R_k)\cdots D^{(l)}(\hat R_1)\,\mathbf{c}(0),    
\end{equation}
where $D^{(l)}(\hat R_p)$ are SU(2) representation matrices. The non-commutativity of the rotations ensures that $\mathbf{c}(0)$ spreads over multiple components for generic initial states. Hence, for some finite $k$, at least two amplitudes become nonzero, implying that $\rho_1^{(N)}$ has rank $\ge 2$ and therefore positive entropy.

However, this spreading leads to entanglement only under additional conditions. This distinction is illustrated in Fig.~\ref{fig5}(c), where coherent states exhibit limited entanglement growth under stroboscopic evolution, whereas squeezed states show rapid enhancement. The stroboscopic evolution implements a passive SU(2) linear transformation of the bosonic modes, $\hat a_1 \rightarrow u_k \hat a_1 + v_k \hat a_2$,
$\hat a_2 \rightarrow -v_k^* \hat a_1 + u_k^* \hat a_2$, with $|u_k|^2+|v_k|^2=1$. Such transformations preserve the class of classical states, namely those with a positive Glauber–Sudarshan $P$-representation. It is well established that passive linear transformations cannot generate entanglement from classical input states \cite{PhysRevA.65.032323,PhysRevA.89.052302}.

Therefore, entanglement generation requires that the initial state be nonclassical. In addition, if the evolution does not populate both modes, the state remains effectively single-mode and no entanglement is produced.

This completes the proof that a necessary condition for entanglement generation is that the initial state is nonclassical and the stroboscopic evolution populates both modes. 
Under these conditions, correlations between the modes develop, and the reduced density matrix becomes mixed, implying two-mode entanglement.

\onecolumngrid
\newpage

\section*{Supplementary Information I: Undisturbed tunneling}

\subsection{Evolution}

The survival probability is defined as

\begin{equation}\label{a11}
P(t)=\left|U(\Omega t ; \theta, 0)\right|^{2}. 
\end{equation}
Equation \eqref{a11} describes periodic evolution with the period $2 \pi / \Omega$. 
In particular, when all the bosons are initially in well 1, i.e., $n_{0}=N, m_{0}=l=N / 2$, we have

\begin{equation}\label{a12}
\left.U(\Omega t ; \theta, \Phi)=[\cos (\Omega t / 2)-i \sin (\Omega t / 2) \cos \theta)\right]^{l}. 
\end{equation}

From Eqs. \eqref{a11} and \eqref{a12}, we obtain the probability of the initial state in the form,

\begin{equation}\label{ap59}
P_{N}(t)=\left(1-\frac{J^{2}}{\Omega^{2}} \sin ^{2} \frac{\Omega t}{2}\right)^{N}. 
\end{equation}

In the resonant case, $\Delta=0$, it becomes,
\begin{equation}\label{ap60}
P_{N}(t)=\cos ^{2 N} \frac{\Omega t}{2}.
\end{equation}

It vanishes when all the bosons are transferred to well 2.
For $\Delta \neq 0$, a complete transfer is impossible. However, for $\Delta^{2} \ll N J^{2}$, the probability of the initial state can decrease from 1 to a very small number. For $N \gg 1$, most of this decay occurs according to the Gaussian law,

\begin{equation}\label{ap61}
P_{N}(t) \approx e^{-N J^{2} t^{2} / 4}, \quad t \ll \Omega^{-1}. 
\end{equation}

There are periodic revivals of the initial state, so that the function \eqref{ap59} includes the equidistant Gaussian peaks $(k=1,2, \ldots)$,

\begin{equation}\label{ap62}
P_{N}(t) \approx e^{-N J^{2}\left(t-t_{2 k}\right)^{2} / 4}, \quad\left|t-t_{2 k}\right| \ll \Omega^{-1}.
\end{equation}

Note that the formulas \eqref{ap61} and \eqref{ap62} are independent of $\Delta$, i.e., the effects of the well detuning are not significant here. In the opposite limit, $\Delta^{2} \gg N J^{2}$, the decay is small,

\begin{equation}\label{ap63}
P_{N}(t) \approx 1-\frac{N J^{2}}{\Delta^{2}} \sin ^{2} \frac{\Delta t}{2} \approx 1.
\end{equation}

Equations \eqref{ap61}-\eqref{ap63} imply that the effective strength of the coupling of the state $|N\rangle$ to the other states is $\sqrt{N} J$. The factor $\sqrt{N}$ here is a cooperative effect.

Let us explicitly consider the initial state \( \ket{\psi(0)} = \ket{l, l} \), corresponding to all bosons in one mode. The m-state amplitudes then evolve as:
\begin{equation}\label{AppendixeqA8}
c_m(t) = \bra{l,m} e^{-i \Omega t L_n} \ket{l,l} = D^{(l)}_{m,l}(0, \theta, \Omega t)= \sum_{m'} d^{(l)}_{m m'}(\theta) \, d^{(l)}_{l m'}(\theta) \, e^{-i m' \Omega t},
\end{equation}
where $D^{(l)}_{m,l}(0, \theta, \Omega t)$ are the Wigner matrix elements that are related through \eqref{AppendixeqA8} to the reduced Wigner matrix elements~\cite{edmonds1996angular,varshalovich1988quantum} 
\begin{equation} \label{a10}
    d_{mm^\prime}^{(l)}(\theta) = \sum_{k} \frac{(-1)^k \sqrt{(l+m)! (l-m)! (l+m^\prime)! (l-m^\prime)!}}{(l+m-k)! (l-m^\prime-k)! k! (k+m^\prime-m)!} \left(\cos \frac{\theta}{2}\right)^{2l+m-m^\prime-2k} \left(\sin \frac{\theta}{2}\right)^{2k+m^\prime-m},
\end{equation}
with the sum over $k$ such that all factorial arguments are non-negative. An arbitrary initial superposition of $|N,0\rangle$ states with amplitudes $A_N$ can be written in the angular momentum basis $|l,m\rangle$ as
\begin{equation}
    |\psi(t)\rangle = e^{-i\hat H t} \sum_N A_N |N, 0\rangle = \sum_{l \in \{0, \frac{1}{2}, 1, \dots\} } \Tilde{A}_{2l}
    \sum_{m=-l}^{l (N)} c_{l,m}(t) |l,m\rangle.
\end{equation}
$\Tilde{A}_{2l}= \langle l,l| \psi(0)\rangle$  is the initial amplitude of the sector where
\begin{align} 
\langle l,l \mid \psi(0)\rangle
= \left\langle l,l \,\middle|\, \sum_{N=0}^{\infty} A_N \, \ket{N,0}\right\rangle 
= \sum_{N=0}^{\infty} A_N \, \langle l,l \mid N,0\rangle 
= \sum_{N=0}^{\infty} A_N \, \delta_{N,2l},
\end{align}
and $c_{l,m} (t)$ denotes the SU(2) rotation within this sector.

\section*{Supplementary Information II: Tunneling under phase jumps}

Since $\hat{L}_{x}^{\prime \prime}(t)$ is the $x$ component of the vector $\hat{\vec{L}}$ rotated by $\phi(t)$ around the $z$ axis, we obtain that
\begin{equation}\label{b11}
\hat{L}_{x}^{\prime \prime}(t)=(-1)^{[t / \tau]} \hat{L}_{x}.
\end{equation}

From \eqref{b11}, we obtain (for $k=0,1, \ldots$ )

\begin{align}\label{b12}
     \hat{U}(t)= \begin{cases}\hat{R}_{+}\left(t-\theta_{2 k}\right) \hat{R}_{-+}^{k}, & \theta_{2 k} \leq t<\theta_{2 k+1},  \\ 
     \hat{R}_{-}\left(t-\theta_{2 k+1}\right) \hat{R}_{+}(\tau) \hat{R}_{-+}^{k}, & \theta_{2 k+1} \leq t<\theta_{2 k+2}.\end{cases}
\end{align}

Here
\begin{align} \nonumber \label{b13}
& \theta_{j}=j \tau, \quad \hat{R}_{-+}=\hat{R}_{-}(\tau) \hat{R}_{+}(\tau),  \\
& \hat{R}_{ \pm}(t)=e^{-i \Omega t \hat{\vec{L}} \cdot \vec{u}_{ \pm}}, \quad \vec{u}_{ \pm}=( \pm \sin \theta, 0, \cos \theta). 
\end{align}

In particular, from \eqref{b12} we have $(k=0,1, \ldots)$

\begin{equation}\label{b14}
    \hat{U}(j \tau)= \begin{cases}\hat{R}_{-+}^{k}, & j=2 k,  \\ \hat{R}_{+}(\tau) \hat{R}_{-+}^{k}, & j=2 k+1.\end{cases}
\end{equation}

To understand the effect of $\hat{R}_{-+}$, we consider the case of a single spin $l=1 / 2$ ($N=1$). Then

\begin{equation}\label{b15}
\hat{R}_{ \pm}(t)=e^{-i(\Omega t / 2) \hat{\sigma} \cdot \vec{u}_{ \pm}}=\cos (\Omega t / 2)-i \sin (\Omega t / 2) \hat{\vec{\sigma}} \cdot \vec{u}_{ \pm}.
\end{equation}

where $\hat{\vec{\sigma}}$ is the vector of the Pauli matrices. Inserting \eqref{b15} into \eqref{b13} and using the Pauli matrices algebra yields
\begin{equation} \label{b16}
\hat{R}_{-+}=e^{-i\left(\alpha_{-+} / 2\right) \hat{\vec{\sigma}} \cdot \vec{u}_{-+}},
\end{equation}
where
\begin{equation}\label{b17}
\alpha_{-+}=2 \arccos \left(1-2 \sin ^{2}(\Omega \tau / 2) \cos ^{2} \theta\right),
\end{equation}
and
\begin{equation}\label{b18}
\vec{u}_{-+}=\frac{(0, \sin (\Omega \tau / 2) \sin \theta, \cos (\Omega \tau / 2))}{\sqrt{1-\sin ^{2}(\Omega \tau / 2) \cos ^{2} \theta}} 
\end{equation}

Thus, $\hat{R}_{-+}$rotates the system by the angle $\alpha_{-+}$around the axis along $\vec{u}_{-+}$. This holds for an arbitrary $L($ or $N)$.

Below we consider the case where
\begin{equation}\label{b19}
\tau=\frac{\pi}{\Omega}.
\end{equation}

Then Eq. \eqref{b16} becomes
\begin{equation}\label{b20}
\hat{R}_{-+}=-e^{2 i \theta \hat{\sigma}_{y}}.
\end{equation}
which means that $\hat{R}_{-+}$rotates the system by the angle $-4 \theta$ around the $y$ axis.

Each phase-flip propagates the wavefunction via the Wigner \( d \)-matrix:
\begin{equation}
    c_m(t) = \sum_{\vec{m}} \prod_{p=1}^{n} d_{m_p m_{p-1}}^{(l)}(\theta_p) e^{-i m_p \Omega \tau},
\end{equation}
where \( \theta_p = \theta \) for even \( p \) and \( \pi - \theta \) for odd \( p \). The modulus becomes:
\begin{equation}
    |c_m(t)|^2 \leq \sum_{\vec{m}} \prod_{p=1}^{n} |d_{m_p m_{p-1}}^{(l)}(\theta_p)|^2.
\end{equation}
For the even numbered $(2k)$ intervals:
\begin{subequations}
\begin{equation}\label{b25}
    c_m(t=2k\tau)= \sum_{\{m_i\}} \left[\prod_{p=1}^{2k} d_{m_p m_{p-1}} (\theta_p) e^{-i m_p \Omega \tau} \right].
\end{equation}

For the odd numbered $(2k+1)$ intervals, we get:
\begin{equation}\label{b24}
    c_m(t=(2k+1)\tau)= \sum_{\{m_i\}} \left[\prod_{p=1}^{2k+1} d_{m_p m_{p-1}} (\theta_p) e^{-i m_p \Omega \tau} \right].
\end{equation}
\end{subequations}
where $m_0 =l$, and $m_{2k+1}=m$, $\theta_p = \theta$ for even $p$, and $\theta_p = \pi-\theta$, for odd $p$.

In the case $n_{0}=N$, we define the functions

\begin{align}\nonumber
& U(2k\tau):=\langle N,0|\hat U(2k\tau)|N,0\rangle=\cos ^{N}(2 k \theta), \\
& U((2k+1)\tau) :=\langle N,0|\hat U((2k+1)\tau)|N,0\rangle=(-i)^{N} \cos ^{N}[(2 k+1) \theta]. 
\end{align}

Then it even- and odd-numbered times, the survival probability is given by
\begin{align}
    P_{n_0}\left(t={2 k \tau}\right)=\left| U(2k\tau)\right|^{2}, \quad P_{n_0}\left(t={(2k+1) \tau}\right)=\left| U((2k+1)\tau)\right|^{2}.
\end{align}

Hence, we obtain
\begin{equation}\label{ap67}
P_{N}\left(t_{j}\right)=\cos ^{2 N}(j \theta) \quad(j=0,1, \ldots). 
\end{equation}

The present control by phase modulation generally increases the degree of tunneling, i.e., decreases the minimum of $P_{N}(t)$. It does not affect $P_{N}(t)$ when $\Delta=0$ and does not significantly affect the rate of decay of $P_{N}(t)$ for $\Delta^{2} \ll N J^{2}$. However, when the tunneling is insignificant in the unperturbed motion, i.e., for $\Delta^{2} \gg N J^{2}$, the control can provide almost complete decay of the initial state. This can be called the strong AZE, since the control changes the decay rate from zero to a final value. In particular, for $\Delta \gg J$, Eq.~\eqref{ap67} yields approximately,

\begin{equation}
P_{N}\left(t_{j}\right)=\exp \left(-j^{2} \frac{N J^{2}}{\Delta^{2}}\right), \quad j \ll \Delta / J.
\end{equation}

\section*{Supplementary Information III: Entanglement Bounds under Modulated and Unmodulated Hamiltonians in Two-Mode Bosonic Systems}

\subsection{Entanglement evaluation and quantification}

To compute two-mode entanglement, we trace out one mode, say mode $2$, and obtain the reduced density matrix for mode $1$:
\begin{equation}\label{a19}
    \rho_1(t)=\text{Tr}_2[\rho (t)],
\end{equation}
where
\begin{equation}\label{a18}
    \rho (t)=|\psi(t)\rangle \langle \psi(t)|= \sum_{m,m^\prime} c_m(t) c_{m^\prime}^*(t) |l,m\rangle \langle l,m^\prime|.
\end{equation}
Since each state $|l, m\rangle$ corresponds to $|n=l+m;N-n=l-m\rangle$, the reduced density matrix is diagonal in the number basis of mode $1$:
\begin{equation}\label{a20}
    \rho_1(t)= \sum_{n=0}^N P(n,t) |n\rangle \langle n|,
\end{equation}
where $|c_m(t)|^2$ is evaluated to be
\begin{equation}\label{a21}
    |c_m(t)|^2 =\sum_{m^\prime,m^{\prime \prime}}  d_{mm^\prime}^{(l)}(\theta)  d_{mm^{\prime \prime}}^{(l)}(\theta) d_{lm^\prime}^{(l)} (\theta) d_{lm^{\prime \prime}}^{(l)} (\theta) e^{-i\Omega (m^{\prime \prime}- m^\prime)t}.
\end{equation}
For the initial product state $|N,0\rangle$ this entropy reads
\begin{equation}\label{a23}
     S(t) = - \sum_{m=-l}^l |c_m(t)|^2 \ln |c_m(t)|^2.
\end{equation}
This entropy grows due to tunneling-induced population mixing. An algebraic manipulation yields the entanglement entropy bound explicitly
\begin{equation} \label{a24}
 \small   S(t) = -\sum_{m=-l}^l \left[ \sum_{m^\prime,m^{\prime \prime}}  d_{mm^\prime}^{(l)}(\theta)  d_{mm^{\prime \prime}}^{(l)}(\theta) d_{lm^\prime}^{(l)} (\theta) d_{lm^{\prime \prime}}^{(l)} (\theta) e^{-i\Omega (m^{\prime \prime}- m^\prime)t} \right] \ln \left[ \sum_{m^\prime,m^{\prime \prime}}  d_{mm^\prime}^{(l)}(\theta)  d_{mm^{\prime \prime}}^{(l)}(\theta) d_{lm^\prime}^{(l)} (\theta) d_{lm^{\prime \prime}}^{(l)} (\theta) e^{-i\Omega (m^{\prime \prime}- m^\prime)t} \right].
\end{equation}

\subsection{Truncated Gaussian Approximation}

According to the Central Limit Theorem, the distribution \( P_m (t)= |c_m(t)|^2 \) evolves into a Gaussian-like profile centered at \( \mu \), with variance \( \sigma^2 \), and truncated beyond the interval \( m \in [-l, l] \):

\begin{equation}
    P_m = \frac{1}{Z} \exp\left(-\frac{(m - \mu)^2}{2\sigma^2} \right), \quad m = -l, \dots, l
\end{equation}

where the normalization factor is:

\begin{equation}
    Z = \sum_{m=-l}^l \exp\left(-\frac{(m - \mu)^2}{2\sigma^2} \right).
\end{equation}

Then the entropy becomes:

\begin{align}
    S(t) &= -\sum_{m=-l}^l P_m \ln P_m, \nonumber\\
         &= \ln Z + \frac{1}{2\sigma^2} \sum_{m=-l}^l P_m (m - \mu)^2, \nonumber\\
         &= \ln Z + \frac{\sigma^2_{\text{eff}}}{2\sigma^2},
\end{align}

where \( \sigma^2_{\text{eff}} = \sum_m P_m (m - \mu)^2 \leq \sigma^2 \). Hence, a simple upper bound on the entropy is:

\begin{equation}
    S(t) \leq \ln Z + \frac{1}{2}.
\end{equation}

We can approximate \( Z \) using the continuous Gaussian integral and the error function of the standard normal distribution.

\subsection{Bounds}

\color{black}

\begin{equation}
    Z \approx \sqrt{2\pi \sigma^2} \cdot \eta(\sigma, l).
\end{equation}

Substituting this normalization into the entropy, we obtain:

\begin{equation}\label{bound1ex}
    S(t) \lesssim \frac{1}{2} \ln (2\pi \sigma^2) + \ln \eta.
\end{equation}
We define the normalized truncation factor:
\begin{equation}
    \eta(\sigma, l) = \Phi\left(\frac{l - \mu}{\sigma}\right) - \Phi\left(\frac{-l - \mu}{\sigma}\right),
\end{equation}
where the standard normal cumulative distribution function (CDF) is
\[
\Phi(x)=\frac{1}{\sqrt{2\pi}}\int_{-\infty}^{x} e^{-t^2/2}\,dt.
\]

The elementary representation of the CDF is
\[
\Phi(x)=\frac{1}{2}\Big[1+\operatorname{erf}\!\big(\tfrac{x}{\sqrt{2}}\big)\Big]
=\frac{1}{2}\,\operatorname{erfc}\!\big(-\tfrac{x}{\sqrt{2}}\big),
\]
where \(\operatorname{erf}(z)=\tfrac{2}{\sqrt{\pi}}\int_0^z e^{-u^2}du\) and \(\operatorname{erfc}(z)=1-\operatorname{erf}(z)\).

We now evaluate this finite-\( l \) entropy bound, accounting for both unmodulated and modulated Hamiltonians.

\subsubsection*{A: Unmodulated Hamiltonian}

For the unmodulated Hamiltonian in

\begin{equation}
    \hat{H} = J L_x + \Delta L_z = \Omega (\cos\theta L_x + \sin\theta L_z), \quad \tan \theta = \frac{J}{\Delta},
\end{equation}
the unmodulated (undisturbed) dynamics is a rotation about the axis \( \vec{n} \), leading to the mean and variance of the resulting distribution:

\[
\mu = l \cos\theta, \quad \sigma^2 \sim l \sin^2\theta.
\]

Then the truncation factor $\eta_{\text{unmod}}$ in \eqref{bound1ex} is 

\begin{equation}
    \eta_{\text{unmod}} = \Phi\left( \frac{l - \mu}{\sqrt{l} \sin\theta} \right) - \Phi\left( \frac{-l - \mu}{\sqrt{l} \sin\theta} \right).
\end{equation}


Using the error function,
\begin{equation}
\eta_{\mathrm{unmod}}
=\tfrac12\Big[
\operatorname{erf}\!\Big(\frac{l-\mu}{\sqrt{2l}\sin\theta}\Big)
-\operatorname{erf}\!\Big(\frac{-l-\mu}{\sqrt{2l}\sin\theta}\Big)
\Big],
\end{equation}

and explicitly substituting \(\mu=l\cos\theta\) we have 
\begin{equation}
    \frac{l-\mu}{\sqrt{l}\sin\theta}
=\sqrt{l}\,\frac{1-\cos\theta}{\sin\theta}
=\sqrt{l}\tan \left(\tfrac{\theta}{2}\right),
\end{equation}

and
\begin{equation}
    \frac{-l-\mu}{\sqrt{l}\sin\theta}
=-\sqrt{l}\,\frac{1+\cos\theta}{\sin\theta}
=-\sqrt{l}\cot \left(\tfrac{\theta}{2}\right).
\end{equation}

Hence, an equivalent form is
\begin{equation}
\eta_{\mathrm{unmod}}
=\tfrac12\left[
\operatorname{erf}\left(\dfrac{\sqrt{l}\,\tan\!\tfrac{\theta}{2}}{\sqrt{2}}\right)
+\operatorname{erf} \left(\dfrac{\sqrt{l}\,\cot\!\tfrac{\theta}{2}}{\sqrt{2}}\right)
\right].
\end{equation}

Hence, the entropy bound becomes:
\begin{equation}
    S_{\text{unmod}}(t) \lesssim \frac{1}{2} \ln (2\pi  l \sin^2 \theta) + \ln \eta_{\text{unmod}}.
\end{equation}

\subsubsection*{B: Modulated (Stroboscopic) Hamiltonian}
Specifically, we apply stroboscopic, periodic $\pi$-phase modulation
The modulated Hamiltonian in the rotating frame is:

\begin{equation}
    \tilde{H}(t) = J (-1)^{\lfloor t/\tau \rfloor} L_x + \Delta L_z.
\end{equation}

Each time step alternates the sign of \( L_x \), leading to a stroboscopic walk in angular momentum space. For \( k \) steps:

\[
\mu \rightarrow 0, \quad \sigma^2 \sim k l \sin^2 \theta.
\]
Then the truncation factor is
\begin{equation}
    \eta_{\text{mod}} = \Phi\left( \frac{l}{\sqrt{k l} \sin\theta} \right) - \Phi\left( \frac{-l}{\sqrt{k l} \sin\theta} \right).
\end{equation}

Equivalently (simplifying \(l/\sqrt{kl}=\sqrt{l/k}\)) and in terms of the erf function, we have:
\begin{equation}
\eta_{\mathrm{mod}}=\operatorname{erf} \left(\dfrac{\sqrt{lk}}{\sqrt{2}\,\sin\theta}\right)
\;=\; \operatorname{erf} \left(\dfrac{1}{\sqrt{2(k/l)}\,\sin\theta}\right).
\end{equation}

Thus, the entropy bound becomes
\begin{equation}
    S_{\text{mod}}(t) \gtrsim \frac{1}{2} \ln (2\pi  k l \sin^2 \theta) + \ln \eta_{\text{mod}}.
\end{equation}


\subsection{Entanglement Enhancement Analysis}

Let us consider the difference in the entanglement entropy

\begin{equation}
    \Delta S = S_{\text{mod}} - S_{\text{unmod}} \gtrsim \frac{1}{2} \ln k + \ln \frac{\eta_{\text{mod}}}{\eta_{\text{unmod}}}.
\end{equation}

The term \(\frac{1}{2} \ln k\) reflects the increased variance due to repeated stroboscopic modulation, and the entropy grows logarithmically with \(k\).
    
The ratio of truncation factors is
\begin{equation}
\frac{\eta_{\text{mod}}}{\eta_{\text{unmod}}} = \frac{\operatorname{erf}\left( \frac{x}{\sqrt{2}} \right)}{\frac{1}{2}\left[\operatorname{erf}\left( \frac{b}{\sqrt{2}} \right) - \operatorname{erf}\left( \frac{a}{\sqrt{2}} \right) \right]}.
\end{equation}
 For typical \(\theta\), \(a\) is large negative, so \(\operatorname{erf}(a/\sqrt{2}) \approx -1\), yielding
\begin{equation}
    \eta_{\text{unmod}} \approx \frac{1}{2} \left[ \operatorname{erf}\left( \frac{b}{\sqrt{2}} \right) + 1 \right],
\end{equation}
which is independent of \(k\) and saturates.
    
In contrast, \(\eta_{\text{mod}} = \operatorname{erf}(x/\sqrt{2})\) approaches 1 as \(k\) increases, since \(x = \sqrt{\frac{l}{k}} \frac{1}{\sin\theta}\) decreases with \(k\). Therefore, the modulated case both increases variance and effectively reduces truncation errors over time, enhancing entropy.

From \eqref{a23} we find
\begin{equation} \label{a24}
    S (t) \lesssim \frac{1}{2} \ln (2\pi l \sin^2 \theta) + \ln \eta, \quad  \eta=\tfrac12\left[
\operatorname{erf} \Big(\frac{l-\mu}{\sqrt{2l}\sin\theta}\Big)
-\operatorname{erf} \Big(\frac{-l-\mu}{\sqrt{2l}\sin\theta}\Big) \right], \quad \mu= l\cos \theta.
\end{equation}

The error function, $\operatorname{erf} (\bullet)$, measures the probability that a normally distributed random variable lies within a certain range around the mean values, $(l-\mu)/\sqrt{2l} \sin \theta$ and $(l+\mu)/\sqrt{2l} \sin \theta$. The difference between the two error entities renders the entanglement entropy small for $\mu << l$, or equivalently for the large-detuning limit $NJ^2 << \Delta^2$.

Let us approximate the discrete probability distribution $P_N(m,t)$ by a truncated Gaussian over the finite domain $m\in[-l,l]$. For a squeezed initial state, the ensemble includes many 
$l$-sectors weighted by $|c_N|^2$. However, because each $l$ evolves independently, the effective population distribution at time $t$ can be represented by a weighted mixture of Gaussian components:
\begin{equation}
    P_{\mathrm{tot}}(m,t) 
    \simeq \frac{1}{\mathcal{N}} 
    \sum_{N\,\text{even}} |c_N|^2
    \exp\!\left[-\frac{(m-\mu_N)^2}{2\sigma_N^2}\right],
\end{equation}
where $\mu_N$ and $\sigma_N^2$ correspond to the 
unmodulated or modulated evolution for each $N$,
and $\mathcal{N}$ ensures normalization within $[-l,l]$.

For large $N$ (and correspondingly large $l$), this mixture 
is itself approximately Gaussian with 
mean $\mu_{\mathrm{tot}}$ and variance $\sigma_{\mathrm{tot}}^2$ given previously.
The entropy of a truncated Gaussian distribution supported on $[-L,L]$ 
is then bounded as:
\begin{equation}
    S(t) = -\sum_m P(m,t)\ln P(m,t)
    \;\approx\;
    \frac{1}{2}\ln(2\pi \,\sigma_{\mathrm{eff}}^2)
    + \ln \eta,
    \label{eq:entropy_bound}
\end{equation}
where $\sigma_{\mathrm{eff}}^2$ is the effective variance 
($\sigma_{\mathrm{tot}}^2$ for uncontrolled or 
$\sigma_{\mathrm{tot},k}^2$ for modulated cases),
and $\eta$ accounts for finite support corrections:
\begin{equation}
    \eta = 
    \frac{1}{
    \sqrt{2\pi}\,\sigma_{\mathrm{eff}}\,
    \mathrm{erf}\!\left(\frac{L-\mu_{\mathrm{eff}}}{\sqrt{2}\sigma_{\mathrm{eff}}}\right)
    } \int_{-L}^{L} 
    e^{-\frac{(m-\mu_{\mathrm{eff}})^2}{2\sigma_{\mathrm{eff}}^2}}\,dm.
\end{equation}

In the uncontrolled case, the squeezed-vacuum entanglement entropy
has nonzero mean $\mu_{\mathrm{tot}}$ and variance $\sigma_{\mathrm{tot}}^2$, 
given by
\begin{align} \nonumber
    \mu_{\mathrm{tot}} &= \frac{\sinh^2 r}{2}\cos\theta,\\[4pt]
    \sigma_{\mathrm{tot}}^2 
    &= \frac{\sinh^2 r}{2}\Big[\sin^2\theta 
      + \cos^2\theta(\sinh^2 r + 1)\Big].
\end{align}
The corresponding entropy bound reads
\begin{equation}
    S_{\mathrm{unmod}}(t)
    \le 
    \frac{1}{2}\ln(2\pi \,\sigma_{\mathrm{tot}}^2)
    + \ln \eta_{\mathrm{unmod}},
    \label{eq:Sunmod_bound}
\end{equation}
with 
\begin{equation}
    \eta_{\mathrm{unmod}}
    = 
    \frac{1}{2}
    \left[
    \mathrm{erf}\!\left(\frac{l-\mu_{\mathrm{tot}}}{\sqrt{2}\sigma_{\mathrm{tot}}}\right)
    +
    \mathrm{erf}\!\left(\frac{l+\mu_{\mathrm{tot}}}{\sqrt{2}\sigma_{\mathrm{tot}}}\right)
    \right].
\end{equation}
When $\sigma_{\mathrm{tot}}\ll l$, $\eta_{\mathrm{unmod}}\approx1$, 
and Eq.~\eqref{eq:Sunmod_bound} simplifies to the continuous Gaussian entropy.

For the periodically modulated Hamiltonian, 
each $N$-sector evolves with zero mean and variance
\[
    \sigma_{N,k}^2 = k\,l\,\sin^2\theta,
\]
and the total mixture has
\[
    \sigma_{\mathrm{tot},k}^2 
    = k\,\frac{\sinh^2 r}{2}\sin^2\theta.
\]
Here, $k$ denotes the stroboscopic index corresponding to the 
$k$-th modulation period, with $t_k =kT$. All quantities are evaluated after $k$ cycles of periodic driving
The corresponding entropy bound is
\begin{equation}
    S_{\mathrm{mod}}(t_k)
    \le 
    \frac{1}{2}\ln(2\pi \,\sigma_{\mathrm{tot},k}^2)
    + \ln \eta_{\mathrm{mod},k},
\end{equation}
where the truncation factor (symmetric around $m=0$) is
\begin{equation}
    \eta_{\mathrm{mod},k}
    = 
    \mathrm{erf}\!\left(\frac{l}{\sqrt{2}\sigma_{\mathrm{tot},k}}\right)
    - \frac{\sqrt{2}\sigma_{\mathrm{tot},k}}{\sqrt{\pi}l}
      e^{-l^2/(2\sigma_{\mathrm{tot},k}^2)}.
\end{equation}
As $k$ increases, $\sigma_{\mathrm{tot},k}\propto\sqrt{k}$, 
so $\eta_{\mathrm{mod},k}\to1$ and the distribution approaches 
a fully delocalized Gaussian over the allowed range.

\textit{Relative entropy enhancement:}
Subtracting the two bounds yields
\begin{equation}
    \Delta S_k 
    = S_{\mathrm{mod}}(t_k) - S_{\mathrm{unmod}}(t)
    \gtrsim 
    \frac{1}{2}\ln
    \left[
        \frac{\sigma_{\mathrm{tot},k}^2}{\sigma_{\mathrm{tot}}^2}
    \right]
    + \ln\!\left(
        \frac{\eta_{\mathrm{mod},k}}{\eta_{\mathrm{unmod}}}
    \right),
\end{equation}
and using the explicit variances gives
\begin{equation}
    \frac{\sigma_{\mathrm{tot},k}^2}{\sigma_{\mathrm{tot}}^2}
    = 
    \frac{
        k\,\sin^2\theta
    }{
        \sin^2\theta
        + \cos^2\theta(\sinh^2 r + 1)
    }.
\end{equation}
Hence,
\begin{equation}
    \Delta S_k 
    \gtrsim 
    \frac{1}{2}\ln k 
    + 
    \frac{1}{2}
    \ln\!\left(
        \frac{
            \sin^2\theta
        }{
            \sin^2\theta + \cos^2\theta(\sinh^2 r + 1)
        }
    \right)
    + 
    \ln\!\left(
        \frac{\eta_{\mathrm{mod},k}}{\eta_{\mathrm{unmod}}}
    \right).
\end{equation}

\section*{Supplementary Information IV: Theorem 1 applications for $N=1$, and $N=2$ spin-coherent and non-coherent states}

Because \(N\) is conserved by the Hamiltonian, the full Hilbert space decomposes as
\begin{equation}
    \mathcal{H} = \bigoplus_{N=0}^\infty \mathcal{H}_N,
\end{equation}
where \(\mathcal{H}_N\) is the subspace with \(N\) bosons shared by the two modes.

An SU(2) coherent (spin coherent) state in the \(N\)-sector can be written in the Fock basis as the
\begin{equation}
    \ket{\theta,\phi}_N
    =
    \sum_{n=0}^N 
    \sqrt{\binom{N}{n}}\,
    \left(\cos\frac{\theta}{2}\right)^n
    \left(e^{i\phi} \sin\frac{\theta}{2}\right)^{N-n}
    \ket{n, N-n}.
    \label{eq:SU2-coherent}
\end{equation}

Consider the stroboscopic evolution over the driving period:
\begin{equation}
    U_{\text{step}} = U_- U_+,
\end{equation}
where \(U_+\) and \(U_-\) are the time-evolution operators for the alternating \(+J\Tilde{L}_x\) and \(-J\Tilde{L}_x\) rotations respectively.

\begin{quote}
If, for some \(N\), the initial state \(\ket{\psi_N(0)}\), is \emph{not} an SU(2) coherent state aligned with a symmetry axis of the control protocol, then after a finite number of periods \(k\), the evolved state \(\ket{\psi_N(k\tau)}\) has support on at least two different basis states \(\ket{\ell,m_1}\) and \(\ket{\ell,m_2}\) with \(m_1 \neq m_2\).
\end{quote}

The SU(2) coherent state \(\ket{\theta,\phi}_N\) defined in Eq.~\eqref{eq:SU2-coherent} corresponds to a total spin $l=N/2$ pointing in the direction
\[
    \bm{n}(\theta,\phi)
    =
    (\sin\theta\cos\phi,\; \sin\theta\sin\phi,\; \cos\theta)
\]
on the sphere hyper. It can be generated from \(\ket{N,0}\)
by a suitable SU(2) rotation to align it with \(\bm{\vec{n}}\).
One period of \(U_{\text{step}}\) induces an SU(2) operation on each \(\mathcal{H}_N\) that can be viewed as a rotation on the hypersphere of angle \(\Phi\) about an axis \(\bm{n}_\text{ctrl}\):
\begin{equation}
    U_{\text{step}}^{(N)} \sim \exp\big(-i \Phi\, \bm{n}_\text{ctrl}\cdot\bm{L}\big).
\end{equation}
The direction \(\bm{n}_\text{ctrl}\) is a \emph{symmetry axis} of the
stroboscopic map: if the spin points exactly along \(\bm{n}_\text{ctrl}\),
then \(U_{\text{step}}^{(N)}\) only rotates it \emph{around} that axis
and leaves the direction of the spin unchanged. The Bloch vector tip stays at a fixed point on sphere; only a global phase may change.

The protocol may have several such symmetry axes (depending on the
symmetries of the driving \(H(t)\)). The key point is that rotations about those axes do
not \textit{spread} the state in the \(\ket{\ell,m}\) basis: a spin coherent state
pointing along \(\bm{n}_\text{ctrl}\) stays coherent and aligned under the map.

For such a state, repeated application of \(U_{\text{step}}^{(N)}\) does \emph{not}
change the direction of \(\langle \bm{L} \rangle\). Consequently, the amplitudes
in the \(\ket{\ell,m}\) basis do not spread the system only accumulates phases.

There are two ways to violate the condition and undergo spreading, i.e. two-mode entanglement under $k$ stroboscopic control steps:
i) \(\ket{\psi_N(0)}\) is not SU(2) coherent at all
    ( any superposition that cannot be written as a single
    rotated \(\ket{N,0}\)); or
    ii) the state is coherent, but its spin
    direction \(\bm{n}(\theta,\phi)\) is \emph{not} one of the symmetry axes
    \(\bm{n}_\text{ctrl}\) of the protocol.
In either case, the state vector under $k$-steps stroboscopic control moves along some nontrivial orbit, which in the \(\ket{\ell,m}\) (or \(\ket{n,N-n}\)) basis manifests as a state
that acquires support on multiple \(m\)-values. This spreading ensures entanglement generation.

\subsection{\(N = 1\): Simplest Case}

For \(N=1\), there is only one boson total. The basis is
\begin{equation}
    \ket{1,0} \equiv \text{particle in mode \(a\)}, \qquad
    \ket{0,1} \equiv \text{particle in mode \(b\)}.
\end{equation}
We define
\begin{equation}
    \ket{\uparrow} = \ket{1,0}, \qquad \ket{\downarrow} = \ket{0,1}.
\end{equation}
Any pure state can be written as
\begin{equation}
    \ket{\psi_1} = \alpha \ket{\uparrow} + \theta \ket{\downarrow},
    \qquad |\alpha|^2 + |\theta|^2 = 1.
\end{equation}
This is a spin-\(\frac12\) state on the Bloch sphere. For spin-\(\frac12\),
\emph{every} pure state is SU(2) coherent; thus, for \(N=1\) the notion of
``coherent vs non-coherent'' is trivial. What matters is whether
\(\ket{\psi_1}\) is aligned with a symmetry axis of \(U_{\text{step}}\).

Take, for concreteness, the initial state as
\begin{equation}
    \ket{\psi_1(0)} = \ket{1,0} = \ket{\uparrow}.
\end{equation}
One stroboscopic step yields
\begin{equation}
    \ket{\psi_1(2\tau)} = U_{\text{step}} \ket{\uparrow}.
\end{equation}
Unless \(\ket{\uparrow}\) is an eigenstate of \(U_{\text{step}}\),
the result is a nontrivial superposition
\begin{equation}
    \ket{\psi_1(2\tau)} = a \ket{1,0} + b \ket{0,1}, \qquad
    a,b \neq 0.
\end{equation}
This state is already entangled in the mode sense: tracing out mode \(b\) gives
\begin{equation}
    \rho_a = \mathrm{Tr}_b \left( \ket{\psi_1(2\tau)}\bra{\psi_1(2\tau)} \right)
           = |a|^2 \ket{1}\bra{1} + |b|^2 \ket{0}\bra{0},
\end{equation}
which is mixed (\(\rho_a^2 \neq \rho_a\)) whenever \(0 < |a| < 1\). Hence the
global pure state \(\ket{\psi_1(2\tau)}\) is entangled across modes \(a\) and \(b\).

This illustrates the mechanism in the simplest sector: unless the initial state
is a very special invariant of the stroboscopic map, the protocol generates
superpositions of different \(\ket{n,N-n}\) components and thus entanglement.

\subsection{ \(N = 2\): Coherent vs Non-Coherent spin states}

Now take \(N=2\). The basis is
\begin{equation}
    \ket{2,0}, \quad \ket{1,1}, \quad \ket{0,2}.
\end{equation}
This corresponds to spin \(\ell=1\) with \(m = 1,0,-1\).

An SU(2) coherent state in this sector has the form
\begin{align} \nonumber
    \ket{\theta,\phi}_2
    &\propto 
    \left(
        \cos\frac{\theta}{2}\, a^\dagger 
        + e^{i\phi} \sin\frac{\theta}{2}\, b^\dagger
    \right)^2 \ket{0,0}, \\
    &=
    \cos^2\frac{\theta}{2} \ket{2,0}
    + \sqrt{2}\, \sin\frac{\theta}{2}\cos\frac{\theta}{2}\, e^{i\phi} \ket{1,1}
    + \sin^2\frac{\theta}{2}\, e^{2i\phi} \ket{0,2}.
\end{align}
All three basis states appear with binomial weights determined by \(\theta,\phi\).

A strongly non-coherent example is the NOON state
\begin{equation}
    \ket{\text{NOON}_2}
    = \frac{1}{\sqrt{2}}\left( \ket{2,0} + \ket{0,2} \right).
\end{equation}
There is no \(\ket{1,1}\) component, so the amplitudes cannot be written in the
binomial form above for any \(\theta,\phi\). Thus \(\ket{\text{NOON}_2}\) is
\emph{not} SU(2) coherent.
Under the stroboscopic protocol, SU(2) coherent states aligned with a symmetry axis
may simply precess without spreading in the \(m\) quantum number. By contrast,
a non-coherent state like \(\ket{\text{NOON}_2}\) has no such rigidity and will
generically evolve into a superposition involving \(\ket{2,0}\), \(\ket{1,1}\),
and \(\ket{0,2}\) after finitely many steps (unless it happens to be an eigenstate
of \(U_{\text{step}}\), which is non-generic).
The presence of at least two populated basis states in \(\mathcal{H}_2\) implies
entanglement between the modes, since the global pure state cannot be written as
a product \(\ket{\phi}_a \otimes \ket{\chi}_b\).

\subsubsection{ Coherent state as a rotated highest-weight state.}

Let \(\ell=1\). The highest-weight state is
\[
    \ket{\ell=1,m=1} \equiv \ket{2,0}.
\]
An SU(2) coherent state pointing in direction
\[
    \bm{n}(\theta,\phi)
    = (\sin\theta\cos\phi,\; \sin\theta\sin\phi,\; \cos\theta)
\]
can be defined as
\begin{equation}
    \ket{\theta,\phi}_2
    = R(\theta,\phi)\,\ket{\ell=1,m=1},
\end{equation}
where, for example,
\begin{equation}
    R(\theta,\phi) = e^{-i\phi L_z}e^{-i\theta L_y},
\end{equation}
is the rotation that takes the \(z\)-axis to the direction \(\bm{n}(\theta,\phi)\).

Define the spin component along \(\bm{n}\) as
\begin{equation}
    L_{\bm{n}}
    := \bm{n}\cdot\bm{L}
    = n_x L_x + n_y L_y + n_z L_z
    = R(\theta,\phi)\,L_z\,R^\dagger(\theta,\phi).
\end{equation}
We then have
\begin{align}\nonumber
    L_{\bm{n}} \ket{\theta,\phi}_2
    &= R(\theta,\phi)\,L_z\,R^\dagger(\theta,\phi)\,
       R(\theta,\phi)\ket{1,1}, \\ \nonumber
    &= R(\theta,\phi)\,L_z\,\ket{1,1}, \\ \nonumber
    &= R(\theta,\phi)\,(+1)\ket{1,1}, \\
    &= +1\,\ket{\theta,\phi}_2.
\end{align}
So for \(N=2\), the coherent state is an eigenstate of \(L_{\bm{n}}\) with eigenvalue
\(\ell=1\).

Now consider a rotation by angle \(\alpha\) about the axis \(\bm{n}\):
\begin{equation}
    U_{\bm{n}}(\alpha) := e^{-i\alpha L_{\bm{n}}}.
\end{equation}
Using the eigenvalue relation,
\begin{align} \nonumber
    U_{\bm{n}}(\alpha)\,\ket{\theta,\phi}_2
    &= e^{-i\alpha L_{\bm{n}}}\,\ket{\theta,\phi}_2, \\  \nonumber
    &= e^{-i\alpha\ell}\,\ket{\theta,\phi}_2, \\  
    &= e^{-i\alpha}\,\ket{\theta,\phi}_2.
\end{align}
Thus, the state is invariant \emph{up to a phase} under any rotation about its own
Bloch direction. The associated projector is strictly invariant:
\begin{equation}
    U_{\bm{n}}(\alpha)\,\ket{\theta,\phi}_2\!\bra{\theta,\phi}_2\,
    U_{\bm{n}}^\dagger(\alpha)
    = \ket{\theta,\phi}_2\!\bra{\theta,\phi}_2.
\end{equation}

\subsubsection{ Explicit check for a rotation about a different axis.}

To see that \(\ket{\theta,\phi}_2\) is \emph{not} invariant under an arbitrary
rotation, let us act with a rotation about the \(z\)-axis:
\begin{equation}
    U_z(\gamma) := e^{-i\gamma L_z}.
\end{equation}
In the basis \(\{\ket{1,1},\ket{1,0},\ket{1,-1}\}\), this acts as
\begin{equation}
    U_z(\gamma)\ket{1,m} = e^{-i\gamma m}\ket{1,m}, \qquad m=1,0,-1.
\end{equation}
Applying \(U_z(\gamma)\) to the coherent state yields
\begin{align}  \nonumber
    U_z(\gamma)\ket{\theta,\phi}_2
    &=
    \cos^2\frac{\theta}{2}\,e^{-i\gamma}\ket{1,1}
    + \sqrt{2}\,\sin\frac{\theta}{2}\cos\frac{\theta}{2}\,e^{i\phi}\,e^{0}\ket{1,0}, \\
    &\quad
    + \sin^2\frac{\theta}{2}\,e^{2i\phi}\,e^{+i\gamma}\ket{1,-1}.
\end{align}

For this to be the \emph{same ray} as \(\ket{\theta,\phi}_2\), we need
\begin{equation}
    U_z(\gamma)\ket{\theta,\phi}_2 = e^{i\chi}\ket{\theta,\phi}_2
\end{equation}
for some global phase \(\chi\). Comparing coefficients of
\(\ket{1,1},\ket{1,0},\ket{1,-1}\), we require:
\begin{align} 
    \cos^2\frac{\theta}{2} e^{-i\gamma}
    &= e^{i\chi} \cos^2\frac{\theta}{2}, \label{eq:cond1} \\
    \sqrt{2}\,\sin\frac{\theta}{2}\cos\frac{\theta}{2} e^{i\phi}
    &= e^{i\chi} \sqrt{2}\,\sin\frac{\theta}{2}\cos\frac{\theta}{2} e^{i\phi}, 
    \label{eq:cond2} \\
    \sin^2\frac{\theta}{2} e^{2i\phi} e^{+i\gamma}
    &= e^{i\chi} \sin^2\frac{\theta}{2} e^{2i\phi}. \label{eq:cond3}
\end{align}
If \(\cos(\theta/2)\neq 0\) and \(\sin(\theta/2)\neq 0\) (generic coherent state with
all three components nonzero), Eqs.~\eqref{eq:cond1}--\eqref{eq:cond3} imply
\begin{align} \nonumber
    e^{-i\gamma} &= e^{i\chi}, \\  \nonumber
    1 &= e^{i\chi}, \\
    e^{+i\gamma} &= e^{i\chi}.
\end{align}
From the middle equation, we get \(e^{i\chi}=1\). Then the first and third require
\begin{equation}
    e^{-i\gamma} = 1, \qquad e^{+i\gamma} = 1
    \;\;\Rightarrow\;\; \gamma = 2\pi k,\quad k\in\mathbb{Z}.
\end{equation}
So for a \emph{generic} coherent state (with \(\theta\neq 0,\pi\)) the only
\(z\)-rotations that leave the state invariant up to phase are the trivial ones
(\(\gamma\) an integer multiple of \(2\pi\), i.e.\ the identity).

There are special cases where some amplitudes vanish:
a) If \(\theta = 0\), then \(\ket{\theta,\phi}_2 = \ket{1,1} = \ket{2,0}\), and any rotation about the \(z\)-axis leaves it invariant up to a phase (it is an eigenstate of \(L_z\) with \(m=1\)).
b) If \(\theta = \pi\), then \(\ket{\theta,\phi}_2 = e^{2i\phi}\ket{1,-1}\), again an eigenstate of \(L_z\) with \(m=-1\).

These are precisely the cases where the Bloch direction \(\bm{n}\) coincides
with the \(+z\) or \(-z\) axis, so rotations about \(z\) are rotations about
\(\bm{n}\). Thus, for \(N=2\), the SU(2) coherent state \(\ket{\theta,\phi}_2\) is invariant (up to a phase) under rotations about its own Bloch direction \(\bm{n}(\theta,\phi)\), since it is an eigenstate of \(L_{\bm{n}}\) with eigenvalue \(\ell=1\), and is \emph{not} invariant under generic SU(2) rotations about other axes, except for trivial identity rotations or special cases where the coherent state itself is aligned with that axis.

\section{Supplementary Information V: Robustness of the CAZE scaling against stochastic errors}

The collective anti-Zeno enhancement obtained in the main text relies on the coherent accumulation of identical SU(2) rotations generated by the ideal stroboscopic propagator
\begin{equation}\label{secVeqn1}
U_0=e^{-iH\tau}e^{-i\pi L_z},
\end{equation}
The survival probability is
\begin{equation}
P_N(k)=\cos^{2N}(k\theta) \simeq e^{-Nk^2\theta^2}.
\end{equation}
We now estimate the effect of imperfect control.


\paragraph*{Noisy Floquet propagator:}
In the case of the realistic implementation, errors in switching intervals and phase jumps are quite likely. Suppose that the switching interval and the phase jump fluctuate independently as
\begin{equation}
\tau_j=\tau+\delta\tau_j,
\qquad
\phi_j=\pi+\epsilon_j,
\end{equation}
where the mean ($1^{\rm st}$ moment) is
\begin{equation}
    \langle\delta\tau_j\rangle=\langle\epsilon_j\rangle = 0,
\end{equation}
and the second moment (the variance) is 
\begin{equation}
    \langle \delta\tau_i \delta\tau_j \rangle = \sigma_\tau^2\delta_{ij},
\qquad
\langle\epsilon_i\epsilon_j \rangle = \sigma_\phi^2\delta_{ij}.
\end{equation}

The propagator \eqref{secVeqn1} for one cycle becomes
\begin{equation}\label{propagtor1}
U_j= e^{-iH(\tau+\delta\tau_j)} e^{-i(\pi+\epsilon_j)L_z}.
\end{equation}
Since the errors are small, we have $\Omega\delta\tau_j\ll1$, $|\epsilon_j|\ll1$, expanding Eq.~\eqref{propagtor1} and considering upto the $2^{\rm nd}$ term, we get
\begin{align}
U_j &=e^{-iH\tau} e^{-i\pi L_z} \left(
1-iH\delta\tau_j -i\epsilon_jL_z \right)
+O(\delta^2,\epsilon^2)
\nonumber\\
&=U_0 \left(1-i\delta K_j \right),
\end{align}
where $\delta K_j= H\delta\tau_j + \epsilon_jL_z$.


Let us consider the noise accumulation over many cycles. The total propagator is 
\begin{equation}
U(k)=\prod_{j=1}^{k} U_j.
\end{equation}
Expanding to first order, we get
\begin{equation}
U(k)= U_0^k - i \sum_{j=1}^{k}
U_0^{k-j} \delta K_j U_0^j + O(\delta^2).
\end{equation}
Because the stochastic fluctuations are unbiased, we have
\begin{equation}
\langle\delta K_j \rangle=0.
\end{equation}
Thus, the first-order correction vanishes,
\begin{equation}
\langle U(k) \rangle = U_0^k + O(\delta^2).
\end{equation}

Hence, the coherent SU(2) rotation responsible for the collective anti-Zeno enhancement remains unchanged to first order. The leading correction, therefore, originates from the second-order cumulant,
\begin{equation}
\Big\langle \delta K_i \delta K_j \Big\rangle = \delta_{ij} \left( H^2\sigma_\tau^2 + L_z^2\sigma_\phi^2 \right).
\end{equation}
Since the errors are uncorrelated, we get
\begin{equation}
\sum_{i,j} \langle \delta K_i \delta K_j
\rangle = k \left(H^2\sigma_\tau^2 +
L_z^2\sigma_\phi^2 \right).
\end{equation}

Therefore, the norm of the accumulated stochastic correction obeys
\begin{equation}
\|\delta U\| = O \left(\sqrt{k}\sqrt{
\Omega^2\sigma_\tau^2 + \sigma_\phi^2} \right),
\end{equation}
whereas the coherent rotation generated by $U_0^k$ grows as
\begin{equation}
\Theta_k= k\theta.
\end{equation}
Consequently,
\begin{equation}
\frac{\|\delta U\|}{\Theta_k}= O\left(\frac{\sqrt{\Omega^2\sigma_\tau^2+\sigma_\phi^2}}{\theta\sqrt{k}}\right).
\end{equation}

The ratio decreases with increasing $k$ and therefore, the coherent evolution dominates, provided
\begin{equation}
(\Omega\sigma_\tau)^2 + (\sigma_\phi)^2 \ll 1.
\end{equation}


\begin{figure}
    \centering
    \includegraphics[width=0.95\linewidth]{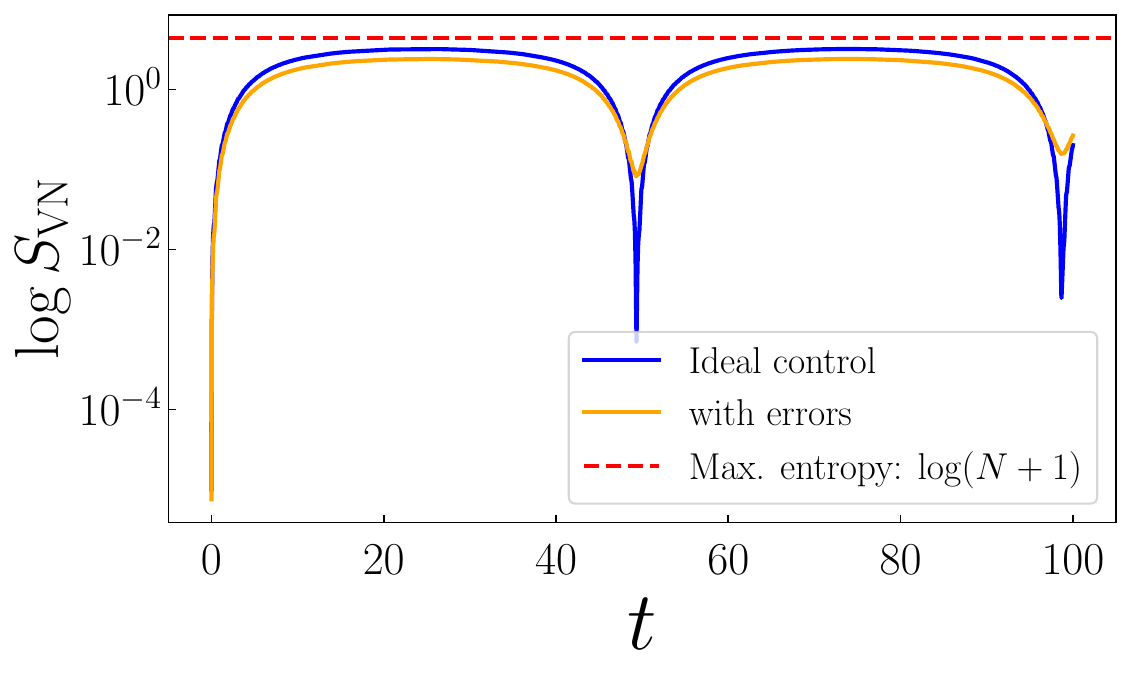}
    \caption{Entanglement entropy as a function of time under phase flips for ideal and erroneous cases with $J=1$, $\Delta =100$, $N=20$, in time units of $\pi/\sqrt{\Delta^2+ J^2}$. The initial state is $\vert N,0\rangle$. The orange solid line shows the entropy averaged over 50 independent stochastic realizations of the control sequence drawn from the Gaussian distribution in the presence of switching-time jitter error about $\sim 0.2\%$ and phase-jump errors about $\sim 0.1\%$.}
    \label{figApp1}
\end{figure}

The quadratic enhancement of the survival exponent,
\begin{equation}
-\ln P_N \propto Nk^2,
\end{equation}
originates from the coherent accumulation of the deterministic SU(2) rotation.
By contrast, stochastic timing jitter and phase calibration errors produce only diffusive broadening of the accumulated rotation, whose variance increases linearly with the number of control cycles.

Hence, the principal effect of realistic control imperfections is not to modify the collective anti-Zeno mechanism itself, but rather to limit the maximum useful number of stroboscopic cycles over which the ideal $k^2N$ scaling remains experimentally observable.

\paragraph*{Finite switching error:} Let us now consider finite switching duration. If the phase update takes a finite time $t_{\rm sw}$, the phase flip remains effectively sudden provided
\begin{equation}
t_{\rm sw}\ll \frac{1}{\Omega}.
\end{equation}
Let us consider that a finite switching time produces an additional error of order
\begin{equation}
\epsilon_{\rm sw}\sim \Omega t_{\rm sw}.
\end{equation}
When timing jitter, phase error, and finite switching error are present, the accumulated control error may be estimated as
\begin{equation}
\epsilon_{\rm tot}(k)\sim \left[(\Omega\sigma_t)^2+(\Omega t_{\rm sw})^2+\sigma_\phi^2 \right],
\end{equation}
The entanglement maxima remain observable as long as
\begin{equation}
\left[(\Omega\sigma_t)^2+(\Omega t_{\rm sw})^2+\sigma_\phi^2\right] \ll 1.
\label{eq:combined_error}
\end{equation}
Equation~\eqref{eq:combined_error} provides an experimental criterion for the robustness of the CAZE protocol. It shows that the protocol is controlled by the dimensionless errors $\Omega\sigma_t$, $\Omega t_{\rm sw}$, and $\sigma_\phi$, rather than by an absolute requirement of infinitely fast switching. Consequently, the entanglement enhancement predicted in the ideal theory remains observable over a finite number of phase-flip operations whenever the accumulated error remains smaller than unity and the total operation time remains below the relevant coherence time (Fig.~\ref{figApp1}).

Thus we can infer that in the ideal case, each stroboscopic step contributes coherently to the accumulated SU(2) rotation angle, so that the effective angle grows linearly with the number of flips, \(\Theta_k \sim k\theta\), and the decay exponent scales as \(N\Theta_k^2 \sim Nk^2\theta^2\), with \(\theta\simeq J/\Delta\) for \(\Delta\gg J\). If the switching interval fluctuates as \(\tau_j=\tau+\delta\tau_j\), or if the phase jump is \(\pi+\epsilon_j\) rather than exactly \(\pi\), the accumulated rotation deviates from the ideal coherent value because each control step contributes a small timing or phase error. These imperfections reduce the constructive interference by introducing small rotation errors at each stroboscopic step. 

Importantly, the protocol does not require infinitely fast switching, but rather that both the switching duration and the timing uncertainty remain small compared with the coherent oscillation period \(2\pi/\Omega\). Assuming the switching error is sufficiently small, each phase update still produces nearly the desired SU(2) rotation, and the accumulated deviation from the ideal trajectory remains small over a finite number of operations. Consequently, the entanglement enhancement remains observable over a finite number of phase-flip operations before accumulated control errors become appreciable. Accordingly, the entanglement maxima remain experimentally observable provided the control errors and the measurement time are both short compared with the coherence time.


\begin{thebibliography}{69}%
\makeatletter
\providecommand \@ifxundefined [1]{%
 \@ifx{#1\undefined}
}%
\providecommand \@ifnum [1]{%
 \ifnum #1\expandafter \@firstoftwo
 \else \expandafter \@secondoftwo
 \fi
}%
\providecommand \@ifx [1]{%
 \ifx #1\expandafter \@firstoftwo
 \else \expandafter \@secondoftwo
 \fi
}%
\providecommand \natexlab [1]{#1}%
\providecommand \enquote  [1]{``#1''}%
\providecommand \bibnamefont  [1]{#1}%
\providecommand \bibfnamefont [1]{#1}%
\providecommand \citenamefont [1]{#1}%
\providecommand \href@noop [0]{\@secondoftwo}%
\providecommand \href [0]{\begingroup \@sanitize@url \@href}%
\providecommand \@href[1]{\@@startlink{#1}\@@href}%
\providecommand \@@href[1]{\endgroup#1\@@endlink}%
\providecommand \@sanitize@url [0]{\catcode `\\12\catcode `\$12\catcode `\&12\catcode `\#12\catcode `\^12\catcode `\_12\catcode `\%12\relax}%
\providecommand \@@startlink[1]{}%
\providecommand \@@endlink[0]{}%
\providecommand \url  [0]{\begingroup\@sanitize@url \@url }%
\providecommand \@url [1]{\endgroup\@href {#1}{\urlprefix }}%
\providecommand \urlprefix  [0]{URL }%
\providecommand \Eprint [0]{\href }%
\providecommand \doibase [0]{http://dx.doi.org/}%
\providecommand \selectlanguage [0]{\@gobble}%
\providecommand \bibinfo  [0]{\@secondoftwo}%
\providecommand \bibfield  [0]{\@secondoftwo}%
\providecommand \translation [1]{[#1]}%
\providecommand \BibitemOpen [0]{}%
\providecommand \bibitemStop [0]{}%
\providecommand \bibitemNoStop [0]{.\EOS\space}%
\providecommand \EOS [0]{\spacefactor3000\relax}%
\providecommand \BibitemShut  [1]{\csname bibitem#1\endcsname}%
\let\auto@bib@innerbib\@empty
\bibitem [{\citenamefont {Gamow}(1928)}]{Gamow1928}%
  \BibitemOpen
  \bibfield  {author} {\bibinfo {author} {\bibfnamefont {G.}~\bibnamefont {Gamow}},\ }\bibfield  {title} {\enquote {\bibinfo {title} {Zur quantentheorie des atomkernes},}\ }\href {\doibase 10.1007/BF01343196} {\bibfield  {journal} {\bibinfo  {journal} {Zeitschrift f{\"u}r Physik}\ }\textbf {\bibinfo {volume} {51}},\ \bibinfo {pages} {204--212} (\bibinfo {year} {1928})}\BibitemShut {NoStop}%
\bibitem [{\citenamefont {Belloni}(2016)}]{belloni2016alpha}%
  \BibitemOpen
  \bibfield  {author} {\bibinfo {author} {\bibfnamefont {Fabio}\ \bibnamefont {Belloni}},\ }\bibfield  {title} {\enquote {\bibinfo {title} {Alpha decay in electron environments of increasing density: from the bare nucleus to compressed matter},}\ }\href {\doibase https://doi.org/10.1140/epja/i2016-16032-3} {\bibfield  {journal} {\bibinfo  {journal} {The European Physical Journal A}\ }\textbf {\bibinfo {volume} {52}},\ \bibinfo {pages} {32} (\bibinfo {year} {2016})}\BibitemShut {NoStop}%
\bibitem [{\citenamefont {Ichimaru}(1993)}]{ichimaru1993nuclear}%
  \BibitemOpen
  \bibfield  {author} {\bibinfo {author} {\bibfnamefont {Setsuo}\ \bibnamefont {Ichimaru}},\ }\bibfield  {title} {\enquote {\bibinfo {title} {Nuclear fusion in dense plasmas},}\ }\href {\doibase 10.1103/RevModPhys.65.255} {\bibfield  {journal} {\bibinfo  {journal} {Reviews of Modern Physics}\ }\textbf {\bibinfo {volume} {65}},\ \bibinfo {pages} {255} (\bibinfo {year} {1993})}\BibitemShut {NoStop}%
\bibitem [{\citenamefont {Jackson}(1957)}]{jackson1957catalysis}%
  \BibitemOpen
  \bibfield  {author} {\bibinfo {author} {\bibfnamefont {John~David}\ \bibnamefont {Jackson}},\ }\bibfield  {title} {\enquote {\bibinfo {title} {Catalysis of nuclear reactions between hydrogen isotopes by $\mu$- mesons},}\ }\href {\doibase 10.1103/PhysRev.106.330} {\bibfield  {journal} {\bibinfo  {journal} {Physical Review}\ }\textbf {\bibinfo {volume} {106}},\ \bibinfo {pages} {330} (\bibinfo {year} {1957})}\BibitemShut {NoStop}%
\bibitem [{\citenamefont {Segal}\ \emph {et~al.}(2006)\citenamefont {Segal}, \citenamefont {Seideman}, \citenamefont {Kurizki},\ and\ \citenamefont {Shapiro}}]{segal2006enhancement}%
  \BibitemOpen
  \bibfield  {author} {\bibinfo {author} {\bibfnamefont {Dvira}\ \bibnamefont {Segal}}, \bibinfo {author} {\bibfnamefont {Tamar}\ \bibnamefont {Seideman}}, \bibinfo {author} {\bibfnamefont {Gershon}\ \bibnamefont {Kurizki}}, \ and\ \bibinfo {author} {\bibfnamefont {Moshe}\ \bibnamefont {Shapiro}},\ }\bibfield  {title} {\enquote {\bibinfo {title} {Enhancement of nuclear tunneling through coulomb-barriers using molecular cages},}\ }\href {\doibase https://doi.org/10.1016/j.cplett.2005.12.076} {\bibfield  {journal} {\bibinfo  {journal} {Chemical physics letters}\ }\textbf {\bibinfo {volume} {420}},\ \bibinfo {pages} {241--244} (\bibinfo {year} {2006})}\BibitemShut {NoStop}%
\bibitem [{\citenamefont {Bardeen}(1961)}]{Bardeen1961}%
  \BibitemOpen
  \bibfield  {author} {\bibinfo {author} {\bibfnamefont {J.}~\bibnamefont {Bardeen}},\ }\bibfield  {title} {\enquote {\bibinfo {title} {Tunnelling from a many-particle point of view},}\ }\href {\doibase 10.1103/PhysRevLett.6.57} {\bibfield  {journal} {\bibinfo  {journal} {Physical Review Letters}\ }\textbf {\bibinfo {volume} {6}},\ \bibinfo {pages} {57--59} (\bibinfo {year} {1961})}\BibitemShut {NoStop}%
\bibitem [{\citenamefont {Josephson}(1962)}]{Josephson1962}%
  \BibitemOpen
  \bibfield  {author} {\bibinfo {author} {\bibfnamefont {B.~D.}\ \bibnamefont {Josephson}},\ }\bibfield  {title} {\enquote {\bibinfo {title} {Possible new effects in superconductive tunnelling},}\ }\href {\doibase 10.1016/0031-9163(62)91369-0} {\bibfield  {journal} {\bibinfo  {journal} {Physics Letters}\ }\textbf {\bibinfo {volume} {1}},\ \bibinfo {pages} {251--253} (\bibinfo {year} {1962})}\BibitemShut {NoStop}%
\bibitem [{\citenamefont {Binnig}\ and\ \citenamefont {Rohrer}(1986)}]{BinnigRohrer1986}%
  \BibitemOpen
  \bibfield  {author} {\bibinfo {author} {\bibfnamefont {G.}~\bibnamefont {Binnig}}\ and\ \bibinfo {author} {\bibfnamefont {H.}~\bibnamefont {Rohrer}},\ }\bibfield  {title} {\enquote {\bibinfo {title} {Scanning tunneling microscopy},}\ }\href {\doibase 10.1147/rd.304.0355} {\bibfield  {journal} {\bibinfo  {journal} {IBM Journal of Research and Development}\ }\textbf {\bibinfo {volume} {30}},\ \bibinfo {pages} {355--369} (\bibinfo {year} {1986})}\BibitemShut {NoStop}%
\bibitem [{\citenamefont {Schawlow}\ and\ \citenamefont {Townes}(1958)}]{SchawlowTownes1958}%
  \BibitemOpen
  \bibfield  {author} {\bibinfo {author} {\bibfnamefont {Arthur~L.}\ \bibnamefont {Schawlow}}\ and\ \bibinfo {author} {\bibfnamefont {Charles~H.}\ \bibnamefont {Townes}},\ }\bibfield  {title} {\enquote {\bibinfo {title} {Infrared and optical masers},}\ }\href {\doibase 10.1103/PhysRev.112.1940} {\bibfield  {journal} {\bibinfo  {journal} {Physical Review}\ }\textbf {\bibinfo {volume} {112}},\ \bibinfo {pages} {1940--1949} (\bibinfo {year} {1958})}\BibitemShut {NoStop}%
\bibitem [{\citenamefont {Barone}\ \emph {et~al.}(2004)\citenamefont {Barone}, \citenamefont {Kurizki},\ and\ \citenamefont {Kofman}}]{PhysRevLett.92.200403}%
  \BibitemOpen
  \bibfield  {author} {\bibinfo {author} {\bibfnamefont {A.}~\bibnamefont {Barone}}, \bibinfo {author} {\bibfnamefont {G.}~\bibnamefont {Kurizki}}, \ and\ \bibinfo {author} {\bibfnamefont {A.~G.}\ \bibnamefont {Kofman}},\ }\bibfield  {title} {\enquote {\bibinfo {title} {Dynamical control of macroscopic quantum tunneling},}\ }\href {\doibase 10.1103/PhysRevLett.92.200403} {\bibfield  {journal} {\bibinfo  {journal} {Phys. Rev. Lett.}\ }\textbf {\bibinfo {volume} {92}},\ \bibinfo {pages} {200403} (\bibinfo {year} {2004})}\BibitemShut {NoStop}%
\bibitem [{\citenamefont {Japha}\ and\ \citenamefont {Kurizki}(1996{\natexlab{a}})}]{PhysRevLett.77.2909}%
  \BibitemOpen
  \bibfield  {author} {\bibinfo {author} {\bibfnamefont {Y.}~\bibnamefont {Japha}}\ and\ \bibinfo {author} {\bibfnamefont {G.}~\bibnamefont {Kurizki}},\ }\bibfield  {title} {\enquote {\bibinfo {title} {Spontaneous emission from tunneling two-level atoms},}\ }\href {\doibase 10.1103/PhysRevLett.77.2909} {\bibfield  {journal} {\bibinfo  {journal} {Phys. Rev. Lett.}\ }\textbf {\bibinfo {volume} {77}},\ \bibinfo {pages} {2909--2912} (\bibinfo {year} {1996}{\natexlab{a}})}\BibitemShut {NoStop}%
\bibitem [{\citenamefont {Japha}\ and\ \citenamefont {Kurizki}(1996{\natexlab{b}})}]{PhysRevA.53.586}%
  \BibitemOpen
  \bibfield  {author} {\bibinfo {author} {\bibfnamefont {Y.}~\bibnamefont {Japha}}\ and\ \bibinfo {author} {\bibfnamefont {G.}~\bibnamefont {Kurizki}},\ }\bibfield  {title} {\enquote {\bibinfo {title} {Superluminal delays of coherent pulses in nondissipative media: A universal mechanism},}\ }\href {\doibase 10.1103/PhysRevA.53.586} {\bibfield  {journal} {\bibinfo  {journal} {Phys. Rev. A}\ }\textbf {\bibinfo {volume} {53}},\ \bibinfo {pages} {586--590} (\bibinfo {year} {1996}{\natexlab{b}})}\BibitemShut {NoStop}%
\bibitem [{\citenamefont {Steinberg}\ \emph {et~al.}(1993)\citenamefont {Steinberg}, \citenamefont {Kwiat},\ and\ \citenamefont {Chiao}}]{PhysRevLett.71.708}%
  \BibitemOpen
  \bibfield  {author} {\bibinfo {author} {\bibfnamefont {A.~M.}\ \bibnamefont {Steinberg}}, \bibinfo {author} {\bibfnamefont {P.~G.}\ \bibnamefont {Kwiat}}, \ and\ \bibinfo {author} {\bibfnamefont {R.~Y.}\ \bibnamefont {Chiao}},\ }\bibfield  {title} {\enquote {\bibinfo {title} {Measurement of the single-photon tunneling time},}\ }\href {\doibase 10.1103/PhysRevLett.71.708} {\bibfield  {journal} {\bibinfo  {journal} {Phys. Rev. Lett.}\ }\textbf {\bibinfo {volume} {71}},\ \bibinfo {pages} {708--711} (\bibinfo {year} {1993})}\BibitemShut {NoStop}%
\bibitem [{\citenamefont {Martinis}\ \emph {et~al.}(1987)\citenamefont {Martinis}, \citenamefont {Devoret},\ and\ \citenamefont {Clarke}}]{PhysRevB.35.4682}%
  \BibitemOpen
  \bibfield  {author} {\bibinfo {author} {\bibfnamefont {John~M.}\ \bibnamefont {Martinis}}, \bibinfo {author} {\bibfnamefont {Michel~H.}\ \bibnamefont {Devoret}}, \ and\ \bibinfo {author} {\bibfnamefont {John}\ \bibnamefont {Clarke}},\ }\bibfield  {title} {\enquote {\bibinfo {title} {Experimental tests for the quantum behavior of a macroscopic degree of freedom: The phase difference across a {J}osephson junction},}\ }\href {\doibase 10.1103/PhysRevB.35.4682} {\bibfield  {journal} {\bibinfo  {journal} {Phys. Rev. B}\ }\textbf {\bibinfo {volume} {35}},\ \bibinfo {pages} {4682--4698} (\bibinfo {year} {1987})}\BibitemShut {NoStop}%
\bibitem [{\citenamefont {Clarke}\ \emph {et~al.}()\citenamefont {Clarke}, \citenamefont {Devoret},\ and\ \citenamefont {Martinis}}]{clarkeprize}%
  \BibitemOpen
  \bibfield  {author} {\bibinfo {author} {\bibfnamefont {John}\ \bibnamefont {Clarke}}, \bibinfo {author} {\bibfnamefont {Michel}\ \bibnamefont {Devoret}}, \ and\ \bibinfo {author} {\bibfnamefont {John}\ \bibnamefont {Martinis}},\ }\bibfield  {title} {\enquote {\bibinfo {title} {The prize at the end of the quantum tunnel},}\ }\href@noop {} {\ }\BibitemShut {NoStop}%
\bibitem [{\citenamefont {Leggett}(1984)}]{Leggett1984}%
  \BibitemOpen
  \bibfield  {author} {\bibinfo {author} {\bibfnamefont {A.~J.}\ \bibnamefont {Leggett}},\ }\bibfield  {title} {\enquote {\bibinfo {title} {Quantum tunneling in the presence of an arbitrary linear dissipation mechanism},}\ }\href {\doibase 10.1103/PhysRevB.30.1208} {\bibfield  {journal} {\bibinfo  {journal} {Physical Review B}\ }\textbf {\bibinfo {volume} {30}},\ \bibinfo {pages} {1208--1218} (\bibinfo {year} {1984})}\BibitemShut {NoStop}%
\bibitem [{\citenamefont {Nielsen}\ and\ \citenamefont {Chuang}(2010)}]{nielsen2010quantum}%
  \BibitemOpen
  \bibfield  {author} {\bibinfo {author} {\bibfnamefont {Michael~A}\ \bibnamefont {Nielsen}}\ and\ \bibinfo {author} {\bibfnamefont {Isaac~L}\ \bibnamefont {Chuang}},\ }\href@noop {} {\emph {\bibinfo {title} {Quantum computation and quantum information}}}\ (\bibinfo  {publisher} {Cambridge university press},\ \bibinfo {year} {2010})\BibitemShut {NoStop}%
\bibitem [{\citenamefont {Kimble}(2008)}]{Kimble2008}%
  \BibitemOpen
  \bibfield  {author} {\bibinfo {author} {\bibfnamefont {H.~J.}\ \bibnamefont {Kimble}},\ }\bibfield  {title} {\enquote {\bibinfo {title} {The quantum internet},}\ }\href {\doibase 10.1038/nature07127} {\bibfield  {journal} {\bibinfo  {journal} {Nature}\ }\textbf {\bibinfo {volume} {453}},\ \bibinfo {pages} {1023--1030} (\bibinfo {year} {2008})}\BibitemShut {NoStop}%
\bibitem [{\citenamefont {Kim}\ \emph {et~al.}(2002)\citenamefont {Kim}, \citenamefont {Son}, \citenamefont {Bu\ifmmode~\check{z}\else \v{z}\fi{}ek},\ and\ \citenamefont {Knight}}]{PhysRevA.65.032323}%
  \BibitemOpen
  \bibfield  {author} {\bibinfo {author} {\bibfnamefont {M.~S.}\ \bibnamefont {Kim}}, \bibinfo {author} {\bibfnamefont {W.}~\bibnamefont {Son}}, \bibinfo {author} {\bibfnamefont {V.}~\bibnamefont {Bu\ifmmode~\check{z}\else \v{z}\fi{}ek}}, \ and\ \bibinfo {author} {\bibfnamefont {P.~L.}\ \bibnamefont {Knight}},\ }\bibfield  {title} {\enquote {\bibinfo {title} {Entanglement by a beam splitter: Nonclassicality as a prerequisite for entanglement},}\ }\href {\doibase 10.1103/PhysRevA.65.032323} {\bibfield  {journal} {\bibinfo  {journal} {Phys. Rev. A}\ }\textbf {\bibinfo {volume} {65}},\ \bibinfo {pages} {032323} (\bibinfo {year} {2002})}\BibitemShut {NoStop}%
\bibitem [{\citenamefont {Vedral}\ \emph {et~al.}(1997)\citenamefont {Vedral}, \citenamefont {Plenio}, \citenamefont {Rippin},\ and\ \citenamefont {Knight}}]{PhysRevLett.78.2275}%
  \BibitemOpen
  \bibfield  {author} {\bibinfo {author} {\bibfnamefont {V.}~\bibnamefont {Vedral}}, \bibinfo {author} {\bibfnamefont {M.~B.}\ \bibnamefont {Plenio}}, \bibinfo {author} {\bibfnamefont {M.~A.}\ \bibnamefont {Rippin}}, \ and\ \bibinfo {author} {\bibfnamefont {P.~L.}\ \bibnamefont {Knight}},\ }\bibfield  {title} {\enquote {\bibinfo {title} {Quantifying entanglement},}\ }\href {\doibase 10.1103/PhysRevLett.78.2275} {\bibfield  {journal} {\bibinfo  {journal} {Phys. Rev. Lett.}\ }\textbf {\bibinfo {volume} {78}},\ \bibinfo {pages} {2275--2279} (\bibinfo {year} {1997})}\BibitemShut {NoStop}%
\bibitem [{\citenamefont {Sur}\ \emph {et~al.}(2025)\citenamefont {Sur}, \citenamefont {Chattopadhyay},\ and\ \citenamefont {Kurizki}}]{sur2025molecular}%
  \BibitemOpen
  \bibfield  {author} {\bibinfo {author} {\bibfnamefont {Saikat}\ \bibnamefont {Sur}}, \bibinfo {author} {\bibfnamefont {Pritam}\ \bibnamefont {Chattopadhyay}}, \ and\ \bibinfo {author} {\bibfnamefont {Gershon}\ \bibnamefont {Kurizki}},\ }\bibfield  {title} {\enquote {\bibinfo {title} {Molecular processes as quantum information resources},}\ }\href {\doibase https://doi.org/10.1063/5.0272970} {\bibfield  {journal} {\bibinfo  {journal} {The Journal of Chemical Physics}\ }\textbf {\bibinfo {volume} {163}} (\bibinfo {year} {2025}),\ https://doi.org/10.1063/5.0272970}\BibitemShut {NoStop}%
\bibitem [{\citenamefont {Horodecki}\ \emph {et~al.}(1998)\citenamefont {Horodecki}, \citenamefont {Horodecki},\ and\ \citenamefont {Horodecki}}]{PhysRevLett.80.5239}%
  \BibitemOpen
  \bibfield  {author} {\bibinfo {author} {\bibfnamefont {Micha\l{}}\ \bibnamefont {Horodecki}}, \bibinfo {author} {\bibfnamefont {Pawe\l{}}\ \bibnamefont {Horodecki}}, \ and\ \bibinfo {author} {\bibfnamefont {Ryszard}\ \bibnamefont {Horodecki}},\ }\bibfield  {title} {\enquote {\bibinfo {title} {Mixed-state entanglement and distillation: Is there a ``bound'' entanglement in nature?}}\ }\href {\doibase 10.1103/PhysRevLett.80.5239} {\bibfield  {journal} {\bibinfo  {journal} {Phys. Rev. Lett.}\ }\textbf {\bibinfo {volume} {80}},\ \bibinfo {pages} {5239--5242} (\bibinfo {year} {1998})}\BibitemShut {NoStop}%
\bibitem [{\citenamefont {Smerzi}\ \emph {et~al.}(1997)\citenamefont {Smerzi}, \citenamefont {Fantoni}, \citenamefont {Giovanazzi},\ and\ \citenamefont {Shenoy}}]{Smerzi1997}%
  \BibitemOpen
  \bibfield  {author} {\bibinfo {author} {\bibfnamefont {A.}~\bibnamefont {Smerzi}}, \bibinfo {author} {\bibfnamefont {S.}~\bibnamefont {Fantoni}}, \bibinfo {author} {\bibfnamefont {S.}~\bibnamefont {Giovanazzi}}, \ and\ \bibinfo {author} {\bibfnamefont {S.~R.}\ \bibnamefont {Shenoy}},\ }\bibfield  {title} {\enquote {\bibinfo {title} {Quantum coherent atomic tunneling between two trapped {B}ose--{E}instein condensates},}\ }\href {\doibase 10.1103/PhysRevLett.79.4950} {\bibfield  {journal} {\bibinfo  {journal} {Phys. Rev. Lett.}\ }\textbf {\bibinfo {volume} {79}},\ \bibinfo {pages} {4950--4953} (\bibinfo {year} {1997})}\BibitemShut {NoStop}%
\bibitem [{\citenamefont {Albiez}\ \emph {et~al.}(2005{\natexlab{a}})\citenamefont {Albiez}, \citenamefont {Gati}, \citenamefont {F{\"o}lling}, \citenamefont {Hunsmann}, \citenamefont {Cristiani},\ and\ \citenamefont {Oberthaler}}]{Albiez2005}%
  \BibitemOpen
  \bibfield  {author} {\bibinfo {author} {\bibfnamefont {M.}~\bibnamefont {Albiez}}, \bibinfo {author} {\bibfnamefont {R.}~\bibnamefont {Gati}}, \bibinfo {author} {\bibfnamefont {J.}~\bibnamefont {F{\"o}lling}}, \bibinfo {author} {\bibfnamefont {S.}~\bibnamefont {Hunsmann}}, \bibinfo {author} {\bibfnamefont {M.}~\bibnamefont {Cristiani}}, \ and\ \bibinfo {author} {\bibfnamefont {M.~K.}\ \bibnamefont {Oberthaler}},\ }\bibfield  {title} {\enquote {\bibinfo {title} {Direct observation of tunneling and nonlinear self-trapping in a single bosonic {J}osephson junction},}\ }\href {\doibase 10.1103/PhysRevLett.95.010402} {\bibfield  {journal} {\bibinfo  {journal} {Phys. Rev. Lett.}\ }\textbf {\bibinfo {volume} {95}},\ \bibinfo {pages} {010402} (\bibinfo {year} {2005}{\natexlab{a}})}\BibitemShut {NoStop}%
\bibitem [{\citenamefont {Dalfovo}\ \emph {et~al.}(1999)\citenamefont {Dalfovo}, \citenamefont {Giorgini}, \citenamefont {Pitaevskii},\ and\ \citenamefont {Stringari}}]{Dalfovo1999}%
  \BibitemOpen
  \bibfield  {author} {\bibinfo {author} {\bibfnamefont {F.}~\bibnamefont {Dalfovo}}, \bibinfo {author} {\bibfnamefont {S.}~\bibnamefont {Giorgini}}, \bibinfo {author} {\bibfnamefont {L.~P.}\ \bibnamefont {Pitaevskii}}, \ and\ \bibinfo {author} {\bibfnamefont {S.}~\bibnamefont {Stringari}},\ }\bibfield  {title} {\enquote {\bibinfo {title} {Theory of {B}ose--{E}instein condensation in trapped gases},}\ }\href {\doibase 10.1103/RevModPhys.71.463} {\bibfield  {journal} {\bibinfo  {journal} {Rev. Mod. Phys.}\ }\textbf {\bibinfo {volume} {71}},\ \bibinfo {pages} {463--512} (\bibinfo {year} {1999})}\BibitemShut {NoStop}%
\bibitem [{\citenamefont {Leonhardt}(1997)}]{leonhardt1997measuring}%
  \BibitemOpen
  \bibfield  {author} {\bibinfo {author} {\bibfnamefont {Ulf}\ \bibnamefont {Leonhardt}},\ }\href@noop {} {\emph {\bibinfo {title} {Measuring the quantum state of light}}},\ Vol.~\bibinfo {volume} {22}\ (\bibinfo  {publisher} {Cambridge university press},\ \bibinfo {year} {1997})\BibitemShut {NoStop}%
\bibitem [{\citenamefont {Boto}\ \emph {et~al.}(2000)\citenamefont {Boto}, \citenamefont {Kok}, \citenamefont {Abrams}, \citenamefont {Braunstein}, \citenamefont {Williams},\ and\ \citenamefont {Dowling}}]{Boto2000}%
  \BibitemOpen
  \bibfield  {author} {\bibinfo {author} {\bibfnamefont {A.~N.}\ \bibnamefont {Boto}}, \bibinfo {author} {\bibfnamefont {P.}~\bibnamefont {Kok}}, \bibinfo {author} {\bibfnamefont {D.~S.}\ \bibnamefont {Abrams}}, \bibinfo {author} {\bibfnamefont {S.~L.}\ \bibnamefont {Braunstein}}, \bibinfo {author} {\bibfnamefont {C.~P.}\ \bibnamefont {Williams}}, \ and\ \bibinfo {author} {\bibfnamefont {J.~P.}\ \bibnamefont {Dowling}},\ }\bibfield  {title} {\enquote {\bibinfo {title} {Quantum interferometric optical lithography: Exploiting entanglement to beat the diffraction limit},}\ }\href {\doibase 10.1103/PhysRevLett.85.2733} {\bibfield  {journal} {\bibinfo  {journal} {Physical Review Letters}\ }\textbf {\bibinfo {volume} {85}},\ \bibinfo {pages} {2733--2736} (\bibinfo {year} {2000})}\BibitemShut {NoStop}%
\bibitem [{\citenamefont {Dowling}(2008)}]{Dowling2008}%
  \BibitemOpen
  \bibfield  {author} {\bibinfo {author} {\bibfnamefont {J.~P.}\ \bibnamefont {Dowling}},\ }\bibfield  {title} {\enquote {\bibinfo {title} {Quantum optical metrology -- the lowdown on high-{N00N} states},}\ }\href {\doibase 10.1080/00107510802091298} {\bibfield  {journal} {\bibinfo  {journal} {Contemporary Physics}\ }\textbf {\bibinfo {volume} {49}},\ \bibinfo {pages} {125--143} (\bibinfo {year} {2008})}\BibitemShut {NoStop}%
\bibitem [{\citenamefont {Schwinger}(1965)}]{Schwinger1952}%
  \BibitemOpen
  \bibfield  {author} {\bibinfo {author} {\bibfnamefont {Julian}\ \bibnamefont {Schwinger}},\ }\bibfield  {title} {\enquote {\bibinfo {title} {On angular momentum},}\ }in\ \href@noop {} {\emph {\bibinfo {booktitle} {Quantum Theory of Angular Momentum}}},\ \bibinfo {editor} {edited by\ \bibinfo {editor} {\bibfnamefont {L.~C.}\ \bibnamefont {Biedenharn}}\ and\ \bibinfo {editor} {\bibfnamefont {H.}~\bibnamefont {Van~Dam}}}\ (\bibinfo  {publisher} {Academic Press},\ \bibinfo {year} {1965})\ pp.\ \bibinfo {pages} {229--279},\ \bibinfo {note} {originally AEC Report NYO-3071 (1952)}\BibitemShut {NoStop}%
\bibitem [{\citenamefont {Besse}\ \emph {et~al.}(2020)\citenamefont {Besse}, \citenamefont {Reuer}, \citenamefont {Collodo}, \citenamefont {Wulff}, \citenamefont {Wernli}, \citenamefont {Copetudo}, \citenamefont {Malz}, \citenamefont {Magnard}, \citenamefont {Akin}, \citenamefont {Gabureac} \emph {et~al.}}]{besse2020realizing}%
  \BibitemOpen
  \bibfield  {author} {\bibinfo {author} {\bibfnamefont {Jean-Claude}\ \bibnamefont {Besse}}, \bibinfo {author} {\bibfnamefont {Kevin}\ \bibnamefont {Reuer}}, \bibinfo {author} {\bibfnamefont {Michele~C}\ \bibnamefont {Collodo}}, \bibinfo {author} {\bibfnamefont {Arne}\ \bibnamefont {Wulff}}, \bibinfo {author} {\bibfnamefont {Lucien}\ \bibnamefont {Wernli}}, \bibinfo {author} {\bibfnamefont {Adrian}\ \bibnamefont {Copetudo}}, \bibinfo {author} {\bibfnamefont {Daniel}\ \bibnamefont {Malz}}, \bibinfo {author} {\bibfnamefont {Paul}\ \bibnamefont {Magnard}}, \bibinfo {author} {\bibfnamefont {Abdulkadir}\ \bibnamefont {Akin}}, \bibinfo {author} {\bibfnamefont {Mihai}\ \bibnamefont {Gabureac}},  \emph {et~al.},\ }\bibfield  {title} {\enquote {\bibinfo {title} {Realizing a deterministic source of multipartite-entangled photonic qubits},}\ }\href {\doibase https://doi.org/10.1038/s41467-020-18635-x} {\bibfield  {journal} {\bibinfo  {journal} {Nature communications}\ }\textbf {\bibinfo {volume} {11}},\ \bibinfo
  {pages} {4877} (\bibinfo {year} {2020})}\BibitemShut {NoStop}%
\bibitem [{\citenamefont {Saavedra}\ \emph {et~al.}(2000)\citenamefont {Saavedra}, \citenamefont {Gheri}, \citenamefont {T\"orm\"a}, \citenamefont {Cirac},\ and\ \citenamefont {Zoller}}]{PhysRevA.61.062311}%
  \BibitemOpen
  \bibfield  {author} {\bibinfo {author} {\bibfnamefont {C.}~\bibnamefont {Saavedra}}, \bibinfo {author} {\bibfnamefont {K.~M.}\ \bibnamefont {Gheri}}, \bibinfo {author} {\bibfnamefont {P.}~\bibnamefont {T\"orm\"a}}, \bibinfo {author} {\bibfnamefont {J.~I.}\ \bibnamefont {Cirac}}, \ and\ \bibinfo {author} {\bibfnamefont {P.}~\bibnamefont {Zoller}},\ }\bibfield  {title} {\enquote {\bibinfo {title} {Controlled source of entangled photonic qubits},}\ }\href {\doibase 10.1103/PhysRevA.61.062311} {\bibfield  {journal} {\bibinfo  {journal} {Phys. Rev. A}\ }\textbf {\bibinfo {volume} {61}},\ \bibinfo {pages} {062311} (\bibinfo {year} {2000})}\BibitemShut {NoStop}%
\bibitem [{\citenamefont {Cirac}\ \emph {et~al.}(1997)\citenamefont {Cirac}, \citenamefont {Zoller}, \citenamefont {Kimble},\ and\ \citenamefont {Mabuchi}}]{PhysRevLett.78.3221}%
  \BibitemOpen
  \bibfield  {author} {\bibinfo {author} {\bibfnamefont {J.~I.}\ \bibnamefont {Cirac}}, \bibinfo {author} {\bibfnamefont {P.}~\bibnamefont {Zoller}}, \bibinfo {author} {\bibfnamefont {H.~J.}\ \bibnamefont {Kimble}}, \ and\ \bibinfo {author} {\bibfnamefont {H.}~\bibnamefont {Mabuchi}},\ }\bibfield  {title} {\enquote {\bibinfo {title} {Quantum state transfer and entanglement distribution among distant nodes in a quantum network},}\ }\href {\doibase 10.1103/PhysRevLett.78.3221} {\bibfield  {journal} {\bibinfo  {journal} {Phys. Rev. Lett.}\ }\textbf {\bibinfo {volume} {78}},\ \bibinfo {pages} {3221--3224} (\bibinfo {year} {1997})}\BibitemShut {NoStop}%
\bibitem [{\citenamefont {Kuang}\ \emph {et~al.}(2025)\citenamefont {Kuang}, \citenamefont {Diekmann}, \citenamefont {Fischer}, \citenamefont {Rotter},\ and\ \citenamefont {Gonzalez-Ballestero}}]{kuang2025perfect}%
  \BibitemOpen
  \bibfield  {author} {\bibinfo {author} {\bibfnamefont {Zeyu}\ \bibnamefont {Kuang}}, \bibinfo {author} {\bibfnamefont {Oliver}\ \bibnamefont {Diekmann}}, \bibinfo {author} {\bibfnamefont {Lorenz}\ \bibnamefont {Fischer}}, \bibinfo {author} {\bibfnamefont {Stefan}\ \bibnamefont {Rotter}}, \ and\ \bibinfo {author} {\bibfnamefont {Carlos}\ \bibnamefont {Gonzalez-Ballestero}},\ }\bibfield  {title} {\enquote {\bibinfo {title} {Perfect quantum state transfer in a dispersion-engineered waveguide},}\ }\href@noop {} {\bibfield  {journal} {\bibinfo  {journal} {arXiv preprint arXiv:2512.20212}\ } (\bibinfo {year} {2025})}\BibitemShut {NoStop}%
\bibitem [{\citenamefont {Kok}\ \emph {et~al.}(2007)\citenamefont {Kok}, \citenamefont {Munro}, \citenamefont {Nemoto}, \citenamefont {Ralph}, \citenamefont {Dowling},\ and\ \citenamefont {Milburn}}]{RevModPhys.79.135}%
  \BibitemOpen
  \bibfield  {author} {\bibinfo {author} {\bibfnamefont {Pieter}\ \bibnamefont {Kok}}, \bibinfo {author} {\bibfnamefont {W.~J.}\ \bibnamefont {Munro}}, \bibinfo {author} {\bibfnamefont {Kae}\ \bibnamefont {Nemoto}}, \bibinfo {author} {\bibfnamefont {T.~C.}\ \bibnamefont {Ralph}}, \bibinfo {author} {\bibfnamefont {Jonathan~P.}\ \bibnamefont {Dowling}}, \ and\ \bibinfo {author} {\bibfnamefont {G.~J.}\ \bibnamefont {Milburn}},\ }\bibfield  {title} {\enquote {\bibinfo {title} {Linear optical quantum computing with photonic qubits},}\ }\href {\doibase 10.1103/RevModPhys.79.135} {\bibfield  {journal} {\bibinfo  {journal} {Rev. Mod. Phys.}\ }\textbf {\bibinfo {volume} {79}},\ \bibinfo {pages} {135--174} (\bibinfo {year} {2007})}\BibitemShut {NoStop}%
\bibitem [{\citenamefont {Lentrodt}\ \emph {et~al.}(2023)\citenamefont {Lentrodt}, \citenamefont {Diekmann}, \citenamefont {Keitel}, \citenamefont {Rotter},\ and\ \citenamefont {Evers}}]{PhysRevLett.130.263602}%
  \BibitemOpen
  \bibfield  {author} {\bibinfo {author} {\bibfnamefont {Dominik}\ \bibnamefont {Lentrodt}}, \bibinfo {author} {\bibfnamefont {Oliver}\ \bibnamefont {Diekmann}}, \bibinfo {author} {\bibfnamefont {Christoph~H.}\ \bibnamefont {Keitel}}, \bibinfo {author} {\bibfnamefont {Stefan}\ \bibnamefont {Rotter}}, \ and\ \bibinfo {author} {\bibfnamefont {J\"org}\ \bibnamefont {Evers}},\ }\bibfield  {title} {\enquote {\bibinfo {title} {Certifying multimode light-matter interaction in lossy resonators},}\ }\href {\doibase 10.1103/PhysRevLett.130.263602} {\bibfield  {journal} {\bibinfo  {journal} {Phys. Rev. Lett.}\ }\textbf {\bibinfo {volume} {130}},\ \bibinfo {pages} {263602} (\bibinfo {year} {2023})}\BibitemShut {NoStop}%
\bibitem [{\citenamefont {Rotter}\ and\ \citenamefont {Gigan}(2017)}]{RevModPhys.89.015005}%
  \BibitemOpen
  \bibfield  {author} {\bibinfo {author} {\bibfnamefont {Stefan}\ \bibnamefont {Rotter}}\ and\ \bibinfo {author} {\bibfnamefont {Sylvain}\ \bibnamefont {Gigan}},\ }\bibfield  {title} {\enquote {\bibinfo {title} {Light fields in complex media: Mesoscopic scattering meets wave control},}\ }\href {\doibase 10.1103/RevModPhys.89.015005} {\bibfield  {journal} {\bibinfo  {journal} {Rev. Mod. Phys.}\ }\textbf {\bibinfo {volume} {89}},\ \bibinfo {pages} {015005} (\bibinfo {year} {2017})}\BibitemShut {NoStop}%
\bibitem [{\citenamefont {Lee}\ \emph {et~al.}(2025)\citenamefont {Lee}, \citenamefont {Omkar}, \citenamefont {Teo}, \citenamefont {Lee}, \citenamefont {Kwon}, \citenamefont {Kim},\ and\ \citenamefont {Jeong}}]{lee2025photonic}%
  \BibitemOpen
  \bibfield  {author} {\bibinfo {author} {\bibfnamefont {Jaehak}\ \bibnamefont {Lee}}, \bibinfo {author} {\bibfnamefont {Srikrishna}\ \bibnamefont {Omkar}}, \bibinfo {author} {\bibfnamefont {Yong~Siah}\ \bibnamefont {Teo}}, \bibinfo {author} {\bibfnamefont {Seok-Hyung}\ \bibnamefont {Lee}}, \bibinfo {author} {\bibfnamefont {Hyukjoon}\ \bibnamefont {Kwon}}, \bibinfo {author} {\bibfnamefont {MS}~\bibnamefont {Kim}}, \ and\ \bibinfo {author} {\bibfnamefont {Hyunseok}\ \bibnamefont {Jeong}},\ }\bibfield  {title} {\enquote {\bibinfo {title} {Photonic hybrid quantum computing},}\ }\href {\doibase https://doi.org/10.1016/j.newton.2025.100359} {\bibfield  {journal} {\bibinfo  {journal} {Newton}\ } (\bibinfo {year} {2025}),\ https://doi.org/10.1016/j.newton.2025.100359}\BibitemShut {NoStop}%
\bibitem [{\citenamefont {Marek}\ \emph {et~al.}(2008)\citenamefont {Marek}, \citenamefont {Jeong},\ and\ \citenamefont {Kim}}]{PhysRevA.78.063811}%
  \BibitemOpen
  \bibfield  {author} {\bibinfo {author} {\bibfnamefont {P.}~\bibnamefont {Marek}}, \bibinfo {author} {\bibfnamefont {H.}~\bibnamefont {Jeong}}, \ and\ \bibinfo {author} {\bibfnamefont {M.~S.}\ \bibnamefont {Kim}},\ }\bibfield  {title} {\enquote {\bibinfo {title} {Generating ``squeezed'' superpositions of coherent states using photon addition and subtraction},}\ }\href {\doibase 10.1103/PhysRevA.78.063811} {\bibfield  {journal} {\bibinfo  {journal} {Phys. Rev. A}\ }\textbf {\bibinfo {volume} {78}},\ \bibinfo {pages} {063811} (\bibinfo {year} {2008})}\BibitemShut {NoStop}%
\bibitem [{\citenamefont {Kim}\ \emph {et~al.}(2005)\citenamefont {Kim}, \citenamefont {Park}, \citenamefont {Knight},\ and\ \citenamefont {Jeong}}]{PhysRevA.71.043805}%
  \BibitemOpen
  \bibfield  {author} {\bibinfo {author} {\bibfnamefont {M.~S.}\ \bibnamefont {Kim}}, \bibinfo {author} {\bibfnamefont {E.}~\bibnamefont {Park}}, \bibinfo {author} {\bibfnamefont {P.~L.}\ \bibnamefont {Knight}}, \ and\ \bibinfo {author} {\bibfnamefont {H.}~\bibnamefont {Jeong}},\ }\bibfield  {title} {\enquote {\bibinfo {title} {Nonclassicality of a photon-subtracted {G}aussian field},}\ }\href {\doibase 10.1103/PhysRevA.71.043805} {\bibfield  {journal} {\bibinfo  {journal} {Phys. Rev. A}\ }\textbf {\bibinfo {volume} {71}},\ \bibinfo {pages} {043805} (\bibinfo {year} {2005})}\BibitemShut {NoStop}%
\bibitem [{\citenamefont {Cook}\ and\ \citenamefont {Shore}(1979)}]{PhysRevA.20.539}%
  \BibitemOpen
  \bibfield  {author} {\bibinfo {author} {\bibfnamefont {Richard~J.}\ \bibnamefont {Cook}}\ and\ \bibinfo {author} {\bibfnamefont {Bruce~W.}\ \bibnamefont {Shore}},\ }\bibfield  {title} {\enquote {\bibinfo {title} {Coherent dynamics of {$N$}-level atoms and molecules. {III.} {A}n analytically soluble periodic case},}\ }\href {\doibase 10.1103/PhysRevA.20.539} {\bibfield  {journal} {\bibinfo  {journal} {Phys. Rev. A}\ }\textbf {\bibinfo {volume} {20}},\ \bibinfo {pages} {539--544} (\bibinfo {year} {1979})}\BibitemShut {NoStop}%
\bibitem [{\citenamefont {Shore}(1990)}]{Shore1990}%
  \BibitemOpen
  \bibfield  {author} {\bibinfo {author} {\bibfnamefont {Bruce~W.}\ \bibnamefont {Shore}},\ }\href@noop {} {\emph {\bibinfo {title} {The Theory of Coherent Atomic Excitation}}}\ (\bibinfo  {publisher} {Wiley},\ \bibinfo {year} {1990})\BibitemShut {NoStop}%
\bibitem [{\citenamefont {Agarwal}(2012)}]{agarwal2012quantum}%
  \BibitemOpen
  \bibfield  {author} {\bibinfo {author} {\bibfnamefont {Girish~S}\ \bibnamefont {Agarwal}},\ }\href@noop {} {\emph {\bibinfo {title} {Quantum optics}}}\ (\bibinfo  {publisher} {Cambridge University Press},\ \bibinfo {year} {2012})\BibitemShut {NoStop}%
\bibitem [{\citenamefont {Varshalovich}\ \emph {et~al.}(1988)\citenamefont {Varshalovich}, \citenamefont {Moskalev},\ and\ \citenamefont {Khersonskii}}]{varshalovich1988quantum}%
  \BibitemOpen
  \bibfield  {author} {\bibinfo {author} {\bibfnamefont {Dmitri{\u\i}~Aleksandrovich}\ \bibnamefont {Varshalovich}}, \bibinfo {author} {\bibfnamefont {Anatol{\"\i}~Nikolaevitch}\ \bibnamefont {Moskalev}}, \ and\ \bibinfo {author} {\bibfnamefont {Valerij~Kel'manovǐc}\ \bibnamefont {Khersonskii}},\ }\href@noop {} {\emph {\bibinfo {title} {Quantum theory of angular momentum}}}\ (\bibinfo  {publisher} {World Scientific},\ \bibinfo {year} {1988})\BibitemShut {NoStop}%
\bibitem [{\citenamefont {Edmonds}(1996)}]{edmonds1996angular}%
  \BibitemOpen
  \bibfield  {author} {\bibinfo {author} {\bibfnamefont {Alan~Robert}\ \bibnamefont {Edmonds}},\ }\href@noop {} {\emph {\bibinfo {title} {Angular momentum in quantum mechanics}}},\ Vol.~\bibinfo {volume} {4}\ (\bibinfo  {publisher} {Princeton university press},\ \bibinfo {year} {1996})\BibitemShut {NoStop}%
\bibitem [{\citenamefont {Kurizki}\ and\ \citenamefont {Kofman}(2022)}]{kurizki2022thermodynamics}%
  \BibitemOpen
  \bibfield  {author} {\bibinfo {author} {\bibfnamefont {Gershon}\ \bibnamefont {Kurizki}}\ and\ \bibinfo {author} {\bibfnamefont {Abraham~G}\ \bibnamefont {Kofman}},\ }\href@noop {} {\emph {\bibinfo {title} {Thermodynamics and control of open quantum systems}}}\ (\bibinfo  {publisher} {Cambridge University Press},\ \bibinfo {year} {2022})\BibitemShut {NoStop}%
\bibitem [{\citenamefont {Kofman}\ and\ \citenamefont {Kurizki}(2001)}]{PhysRevLett.87.270405}%
  \BibitemOpen
  \bibfield  {author} {\bibinfo {author} {\bibfnamefont {A.~G.}\ \bibnamefont {Kofman}}\ and\ \bibinfo {author} {\bibfnamefont {G.}~\bibnamefont {Kurizki}},\ }\bibfield  {title} {\enquote {\bibinfo {title} {Universal dynamical control of quantum mechanical decay: Modulation of the coupling to the continuum},}\ }\href {\doibase 10.1103/PhysRevLett.87.270405} {\bibfield  {journal} {\bibinfo  {journal} {Phys. Rev. Lett.}\ }\textbf {\bibinfo {volume} {87}},\ \bibinfo {pages} {270405} (\bibinfo {year} {2001})}\BibitemShut {NoStop}%
\bibitem [{\citenamefont {Kofman}\ and\ \citenamefont {Kurizki}(2000)}]{kofman2000acceleration}%
  \BibitemOpen
  \bibfield  {author} {\bibinfo {author} {\bibfnamefont {AG}~\bibnamefont {Kofman}}\ and\ \bibinfo {author} {\bibfnamefont {Gershon}\ \bibnamefont {Kurizki}},\ }\bibfield  {title} {\enquote {\bibinfo {title} {Acceleration of quantum decay processes by frequent observations},}\ }\href {\doibase https://doi.org/10.1038/35014537} {\bibfield  {journal} {\bibinfo  {journal} {Nature}\ }\textbf {\bibinfo {volume} {405}},\ \bibinfo {pages} {546--550} (\bibinfo {year} {2000})}\BibitemShut {NoStop}%
\bibitem [{\citenamefont {Biagioni}\ \emph {et~al.}(2008)\citenamefont {Biagioni}, \citenamefont {Valle}, \citenamefont {Ornigotti}, \citenamefont {Finazzi}, \citenamefont {Duo}, \citenamefont {Laporta},\ and\ \citenamefont {Longhi}}]{biagioni2008experimental}%
  \BibitemOpen
  \bibfield  {author} {\bibinfo {author} {\bibfnamefont {Paolo}\ \bibnamefont {Biagioni}}, \bibinfo {author} {\bibfnamefont {G~Della}\ \bibnamefont {Valle}}, \bibinfo {author} {\bibfnamefont {Marco}\ \bibnamefont {Ornigotti}}, \bibinfo {author} {\bibfnamefont {Marco}\ \bibnamefont {Finazzi}}, \bibinfo {author} {\bibfnamefont {Lamberto}\ \bibnamefont {Duo}}, \bibinfo {author} {\bibfnamefont {Paolo}\ \bibnamefont {Laporta}}, \ and\ \bibinfo {author} {\bibfnamefont {Stefano}\ \bibnamefont {Longhi}},\ }\bibfield  {title} {\enquote {\bibinfo {title} {Experimental demonstration of the optical {Z}eno effect by scanning tunneling optical microscopy},}\ }\href {\doibase https://doi.org/10.1364/OE.16.003762} {\bibfield  {journal} {\bibinfo  {journal} {Optics Express}\ }\textbf {\bibinfo {volume} {16}},\ \bibinfo {pages} {3762--3767} (\bibinfo {year} {2008})}\BibitemShut {NoStop}%
\bibitem [{\citenamefont {Horodecki}\ \emph {et~al.}(2009)\citenamefont {Horodecki}, \citenamefont {Horodecki}, \citenamefont {Horodecki},\ and\ \citenamefont {Horodecki}}]{RevModPhys.81.865}%
  \BibitemOpen
  \bibfield  {author} {\bibinfo {author} {\bibfnamefont {Ryszard}\ \bibnamefont {Horodecki}}, \bibinfo {author} {\bibfnamefont {Pawe\l{}}\ \bibnamefont {Horodecki}}, \bibinfo {author} {\bibfnamefont {Micha\l{}}\ \bibnamefont {Horodecki}}, \ and\ \bibinfo {author} {\bibfnamefont {Karol}\ \bibnamefont {Horodecki}},\ }\bibfield  {title} {\enquote {\bibinfo {title} {Quantum entanglement},}\ }\href {\doibase 10.1103/RevModPhys.81.865} {\bibfield  {journal} {\bibinfo  {journal} {Rev. Mod. Phys.}\ }\textbf {\bibinfo {volume} {81}},\ \bibinfo {pages} {865--942} (\bibinfo {year} {2009})}\BibitemShut {NoStop}%
\bibitem [{\citenamefont {Gondret}\ \emph {et~al.}(2025)\citenamefont {Gondret}, \citenamefont {Lamirault}, \citenamefont {Dias}, \citenamefont {Leprince}, \citenamefont {Westbrook}, \citenamefont {Cl\'ement},\ and\ \citenamefont {Boiron}}]{1y1pzqhh}%
  \BibitemOpen
  \bibfield  {author} {\bibinfo {author} {\bibfnamefont {Victor}\ \bibnamefont {Gondret}}, \bibinfo {author} {\bibfnamefont {Clothilde}\ \bibnamefont {Lamirault}}, \bibinfo {author} {\bibfnamefont {Rui}\ \bibnamefont {Dias}}, \bibinfo {author} {\bibfnamefont {Charlie}\ \bibnamefont {Leprince}}, \bibinfo {author} {\bibfnamefont {Christoph~I.}\ \bibnamefont {Westbrook}}, \bibinfo {author} {\bibfnamefont {David}\ \bibnamefont {Cl\'ement}}, \ and\ \bibinfo {author} {\bibfnamefont {Denis}\ \bibnamefont {Boiron}},\ }\bibfield  {title} {\enquote {\bibinfo {title} {Quantifying two-mode entanglement of bosonic {G}aussian states from their full counting statistics},}\ }\href {\doibase 10.1103/1y1p-zqhh} {\bibfield  {journal} {\bibinfo  {journal} {Phys. Rev. Lett.}\ }\textbf {\bibinfo {volume} {135}},\ \bibinfo {pages} {100201} (\bibinfo {year} {2025})}\BibitemShut {NoStop}%
\bibitem [{\citenamefont {Plenio}(2005)}]{PhysRevLett.95.090503}%
  \BibitemOpen
  \bibfield  {author} {\bibinfo {author} {\bibfnamefont {M.~B.}\ \bibnamefont {Plenio}},\ }\bibfield  {title} {\enquote {\bibinfo {title} {Logarithmic negativity: A full entanglement monotone that is not convex},}\ }\href {\doibase 10.1103/PhysRevLett.95.090503} {\bibfield  {journal} {\bibinfo  {journal} {Phys. Rev. Lett.}\ }\textbf {\bibinfo {volume} {95}},\ \bibinfo {pages} {090503} (\bibinfo {year} {2005})}\BibitemShut {NoStop}%
\bibitem [{\citenamefont {Adesso}\ and\ \citenamefont {Illuminati}(2007)}]{adesso2007entanglement}%
  \BibitemOpen
  \bibfield  {author} {\bibinfo {author} {\bibfnamefont {Gerardo}\ \bibnamefont {Adesso}}\ and\ \bibinfo {author} {\bibfnamefont {Fabrizio}\ \bibnamefont {Illuminati}},\ }\bibfield  {title} {\enquote {\bibinfo {title} {Entanglement in continuous-variable systems: recent advances and current perspectives},}\ }\href {\doibase 10.1088/1751-8113/40/28/S01} {\bibfield  {journal} {\bibinfo  {journal} {Journal of Physics A: Mathematical and Theoretical}\ }\textbf {\bibinfo {volume} {40}},\ \bibinfo {pages} {7821} (\bibinfo {year} {2007})}\BibitemShut {NoStop}%
\bibitem [{\citenamefont {Vidal}\ and\ \citenamefont {Werner}(2002)}]{PhysRevA.65.032314}%
  \BibitemOpen
  \bibfield  {author} {\bibinfo {author} {\bibfnamefont {G.}~\bibnamefont {Vidal}}\ and\ \bibinfo {author} {\bibfnamefont {R.~F.}\ \bibnamefont {Werner}},\ }\bibfield  {title} {\enquote {\bibinfo {title} {Computable measure of entanglement},}\ }\href {\doibase 10.1103/PhysRevA.65.032314} {\bibfield  {journal} {\bibinfo  {journal} {Phys. Rev. A}\ }\textbf {\bibinfo {volume} {65}},\ \bibinfo {pages} {032314} (\bibinfo {year} {2002})}\BibitemShut {NoStop}%
\bibitem [{\citenamefont {S{\o}rensen}\ \emph {et~al.}(2001)\citenamefont {S{\o}rensen}, \citenamefont {Duan}, \citenamefont {Cirac},\ and\ \citenamefont {Zoller}}]{Sorensen2001}%
  \BibitemOpen
  \bibfield  {author} {\bibinfo {author} {\bibfnamefont {A.~S.}\ \bibnamefont {S{\o}rensen}}, \bibinfo {author} {\bibfnamefont {L.~M.}\ \bibnamefont {Duan}}, \bibinfo {author} {\bibfnamefont {J.~I.}\ \bibnamefont {Cirac}}, \ and\ \bibinfo {author} {\bibfnamefont {P.}~\bibnamefont {Zoller}},\ }\bibfield  {title} {\enquote {\bibinfo {title} {Many-particle entanglement with {B}ose--{E}instein condensates},}\ }\href {\doibase 10.1038/35051038} {\bibfield  {journal} {\bibinfo  {journal} {Nature}\ }\textbf {\bibinfo {volume} {409}},\ \bibinfo {pages} {63--66} (\bibinfo {year} {2001})}\BibitemShut {NoStop}%
\bibitem [{\citenamefont {Chou}\ \emph {et~al.}(2005)\citenamefont {Chou}, \citenamefont {Polyakov}, \citenamefont {Kuzmich},\ and\ \citenamefont {Kimble}}]{Chou2005}%
  \BibitemOpen
  \bibfield  {author} {\bibinfo {author} {\bibfnamefont {C.~W.}\ \bibnamefont {Chou}}, \bibinfo {author} {\bibfnamefont {S.~V.}\ \bibnamefont {Polyakov}}, \bibinfo {author} {\bibfnamefont {A.}~\bibnamefont {Kuzmich}}, \ and\ \bibinfo {author} {\bibfnamefont {H.~J.}\ \bibnamefont {Kimble}},\ }\bibfield  {title} {\enquote {\bibinfo {title} {Measurement-induced entanglement for excitation stored in remote atomic ensembles},}\ }\href {\doibase 10.1038/nature04353} {\bibfield  {journal} {\bibinfo  {journal} {Nature}\ }\textbf {\bibinfo {volume} {438}},\ \bibinfo {pages} {828--832} (\bibinfo {year} {2005})}\BibitemShut {NoStop}%
\bibitem [{\citenamefont {Sherson}\ \emph {et~al.}(2006)\citenamefont {Sherson}, \citenamefont {Krauter}, \citenamefont {Olsson}, \citenamefont {Julsgaard}, \citenamefont {Hammerer}, \citenamefont {Cirac},\ and\ \citenamefont {Polzik}}]{sherson2006quantum}%
  \BibitemOpen
  \bibfield  {author} {\bibinfo {author} {\bibfnamefont {Jacob~F}\ \bibnamefont {Sherson}}, \bibinfo {author} {\bibfnamefont {Hanna}\ \bibnamefont {Krauter}}, \bibinfo {author} {\bibfnamefont {Rasmus~K}\ \bibnamefont {Olsson}}, \bibinfo {author} {\bibfnamefont {Brian}\ \bibnamefont {Julsgaard}}, \bibinfo {author} {\bibfnamefont {Klemens}\ \bibnamefont {Hammerer}}, \bibinfo {author} {\bibfnamefont {Ignacio}\ \bibnamefont {Cirac}}, \ and\ \bibinfo {author} {\bibfnamefont {Eugene~S}\ \bibnamefont {Polzik}},\ }\bibfield  {title} {\enquote {\bibinfo {title} {Quantum teleportation between light and matter},}\ }\href {\doibase https://doi.org/10.1038/nature05136} {\bibfield  {journal} {\bibinfo  {journal} {Nature}\ }\textbf {\bibinfo {volume} {443}},\ \bibinfo {pages} {557--560} (\bibinfo {year} {2006})}\BibitemShut {NoStop}%
\bibitem [{\citenamefont {Snyder}\ and\ \citenamefont {Love}(1983)}]{Snyder1983}%
  \BibitemOpen
  \bibfield  {author} {\bibinfo {author} {\bibfnamefont {A.~W.}\ \bibnamefont {Snyder}}\ and\ \bibinfo {author} {\bibfnamefont {J.~D.}\ \bibnamefont {Love}},\ }\href@noop {} {\emph {\bibinfo {title} {Optical Waveguide Theory}}}\ (\bibinfo  {publisher} {Chapman and Hall},\ \bibinfo {year} {1983})\BibitemShut {NoStop}%
\bibitem [{\citenamefont {Yariv}(1973)}]{Yariv1973}%
  \BibitemOpen
  \bibfield  {author} {\bibinfo {author} {\bibfnamefont {A.}~\bibnamefont {Yariv}},\ }\bibfield  {title} {\enquote {\bibinfo {title} {Coupled-mode theory for guided-wave optics},}\ }\href {\doibase 10.1109/JQE.1973.1077767} {\bibfield  {journal} {\bibinfo  {journal} {IEEE Journal of Quantum Electronics}\ }\textbf {\bibinfo {volume} {9}},\ \bibinfo {pages} {919--933} (\bibinfo {year} {1973})}\BibitemShut {NoStop}%
\bibitem [{\citenamefont {Khodorkovsky}\ \emph {et~al.}(2008)\citenamefont {Khodorkovsky}, \citenamefont {Kurizki},\ and\ \citenamefont {Vardi}}]{PhysRevLett.100.220403}%
  \BibitemOpen
  \bibfield  {author} {\bibinfo {author} {\bibfnamefont {Y.}~\bibnamefont {Khodorkovsky}}, \bibinfo {author} {\bibfnamefont {G.}~\bibnamefont {Kurizki}}, \ and\ \bibinfo {author} {\bibfnamefont {A.}~\bibnamefont {Vardi}},\ }\bibfield  {title} {\enquote {\bibinfo {title} {Bosonic amplification of noise-induced suppression of phase diffusion},}\ }\href {\doibase 10.1103/PhysRevLett.100.220403} {\bibfield  {journal} {\bibinfo  {journal} {Phys. Rev. Lett.}\ }\textbf {\bibinfo {volume} {100}},\ \bibinfo {pages} {220403} (\bibinfo {year} {2008})}\BibitemShut {NoStop}%
\bibitem [{\citenamefont {Albiez}\ \emph {et~al.}(2005{\natexlab{b}})\citenamefont {Albiez}, \citenamefont {Gati}, \citenamefont {F\"olling}, \citenamefont {Hunsmann}, \citenamefont {Cristiani},\ and\ \citenamefont {Oberthaler}}]{PhysRevLett.95.010402}%
  \BibitemOpen
  \bibfield  {author} {\bibinfo {author} {\bibfnamefont {Michael}\ \bibnamefont {Albiez}}, \bibinfo {author} {\bibfnamefont {Rudolf}\ \bibnamefont {Gati}}, \bibinfo {author} {\bibfnamefont {Jonas}\ \bibnamefont {F\"olling}}, \bibinfo {author} {\bibfnamefont {Stefan}\ \bibnamefont {Hunsmann}}, \bibinfo {author} {\bibfnamefont {Matteo}\ \bibnamefont {Cristiani}}, \ and\ \bibinfo {author} {\bibfnamefont {Markus~K.}\ \bibnamefont {Oberthaler}},\ }\bibfield  {title} {\enquote {\bibinfo {title} {Direct observation of tunneling and nonlinear self-trapping in a single bosonic {J}osephson junction},}\ }\href {\doibase 10.1103/PhysRevLett.95.010402} {\bibfield  {journal} {\bibinfo  {journal} {Phys. Rev. Lett.}\ }\textbf {\bibinfo {volume} {95}},\ \bibinfo {pages} {010402} (\bibinfo {year} {2005}{\natexlab{b}})}\BibitemShut {NoStop}%
\bibitem [{\citenamefont {Chapman}\ \emph {et~al.}(2025)\citenamefont {Chapman}, \citenamefont {Kuttner}, \citenamefont {Kellner}, \citenamefont {Sabatti}, \citenamefont {Maeder}, \citenamefont {Finco}, \citenamefont {Kaufmann},\ and\ \citenamefont {Grange}}]{n2y32bmz}%
  \BibitemOpen
  \bibfield  {author} {\bibinfo {author} {\bibfnamefont {Robert~J.}\ \bibnamefont {Chapman}}, \bibinfo {author} {\bibfnamefont {Tristan}\ \bibnamefont {Kuttner}}, \bibinfo {author} {\bibfnamefont {Jost}\ \bibnamefont {Kellner}}, \bibinfo {author} {\bibfnamefont {Alessandra}\ \bibnamefont {Sabatti}}, \bibinfo {author} {\bibfnamefont {Andreas}\ \bibnamefont {Maeder}}, \bibinfo {author} {\bibfnamefont {Giovanni}\ \bibnamefont {Finco}}, \bibinfo {author} {\bibfnamefont {Fabian}\ \bibnamefont {Kaufmann}}, \ and\ \bibinfo {author} {\bibfnamefont {Rachel}\ \bibnamefont {Grange}},\ }\bibfield  {title} {\enquote {\bibinfo {title} {On-chip quantum interference between independent lithium niobate-on-insulator photon-pair sources},}\ }\href {\doibase 10.1103/n2y3-2bmz} {\bibfield  {journal} {\bibinfo  {journal} {Phys. Rev. Lett.}\ }\textbf {\bibinfo {volume} {134}},\ \bibinfo {pages} {223602} (\bibinfo {year} {2025})}\BibitemShut {NoStop}%
\bibitem [{\citenamefont {Taghavi}\ \emph {et~al.}(2025)\citenamefont {Taghavi}, \citenamefont {Esmaeeli}, \citenamefont {Chowdhury}, \citenamefont {Awan}, \citenamefont {Hammood}, \citenamefont {Mitchell}, \citenamefont {Witt}, \citenamefont {Pecinovsky}, \citenamefont {Sickler}, \citenamefont {Young} \emph {et~al.}}]{taghavi2025ghz}%
  \BibitemOpen
  \bibfield  {author} {\bibinfo {author} {\bibfnamefont {Iman}\ \bibnamefont {Taghavi}}, \bibinfo {author} {\bibfnamefont {Omid}\ \bibnamefont {Esmaeeli}}, \bibinfo {author} {\bibfnamefont {Sheri~Jahan}\ \bibnamefont {Chowdhury}}, \bibinfo {author} {\bibfnamefont {Kashif~Masud}\ \bibnamefont {Awan}}, \bibinfo {author} {\bibfnamefont {Mustafa}\ \bibnamefont {Hammood}}, \bibinfo {author} {\bibfnamefont {Matthew}\ \bibnamefont {Mitchell}}, \bibinfo {author} {\bibfnamefont {Donald}\ \bibnamefont {Witt}}, \bibinfo {author} {\bibfnamefont {Cory}\ \bibnamefont {Pecinovsky}}, \bibinfo {author} {\bibfnamefont {Jason}\ \bibnamefont {Sickler}}, \bibinfo {author} {\bibfnamefont {Jeff~F}\ \bibnamefont {Young}},  \emph {et~al.},\ }\bibfield  {title} {\enquote {\bibinfo {title} {{GH}z-rate optical phase shift in light-matter interaction-engineered, silicon-ferroelectric nematic liquid crystals},}\ }\href {\doibase https://doi.org/10.1038/s41467-025-63924-y} {\bibfield  {journal} {\bibinfo  {journal} {Nature Communications}\
  }\textbf {\bibinfo {volume} {16}},\ \bibinfo {pages} {8902} (\bibinfo {year} {2025})}\BibitemShut {NoStop}%
\bibitem [{\citenamefont {Ciccarello}\ and\ \citenamefont {Schneble}(2024)}]{Ciccarello2024}%
  \BibitemOpen
  \bibfield  {author} {\bibinfo {author} {\bibfnamefont {Francesco}\ \bibnamefont {Ciccarello}}\ and\ \bibinfo {author} {\bibfnamefont {D.}~\bibnamefont {Schneble}},\ }\bibfield  {title} {\enquote {\bibinfo {title} {Waveguide quantum electrodynamics},}\ }\href {\doibase 10.1364/OPN.35.1.000034} {\bibfield  {journal} {\bibinfo  {journal} {Optics \& Photonics News}\ }\textbf {\bibinfo {volume} {35}},\ \bibinfo {pages} {34--39} (\bibinfo {year} {2024})}\BibitemShut {NoStop}%
\bibitem [{\citenamefont {Levy-Yeyati}\ \emph {et~al.}(2025)\citenamefont {Levy-Yeyati}, \citenamefont {Vega}, \citenamefont {Ramos},\ and\ \citenamefont {Gonz\'alez-Tudela}}]{PRXQuantum.6.010342}%
  \BibitemOpen
  \bibfield  {author} {\bibinfo {author} {\bibfnamefont {Tom\'as}\ \bibnamefont {Levy-Yeyati}}, \bibinfo {author} {\bibfnamefont {Carlos}\ \bibnamefont {Vega}}, \bibinfo {author} {\bibfnamefont {Tom\'as}\ \bibnamefont {Ramos}}, \ and\ \bibinfo {author} {\bibfnamefont {Alejandro}\ \bibnamefont {Gonz\'alez-Tudela}},\ }\bibfield  {title} {\enquote {\bibinfo {title} {Passive photonic {CZ} gate with two-level emitters in chiral multimode waveguide {QED}},}\ }\href {\doibase 10.1103/PRXQuantum.6.010342} {\bibfield  {journal} {\bibinfo  {journal} {PRX Quantum}\ }\textbf {\bibinfo {volume} {6}},\ \bibinfo {pages} {010342} (\bibinfo {year} {2025})}\BibitemShut {NoStop}%
\bibitem [{\citenamefont {Rosen}\ \emph {et~al.}(2012)\citenamefont {Rosen}, \citenamefont {Afek}, \citenamefont {Israel}, \citenamefont {Ambar},\ and\ \citenamefont {Silberberg}}]{PhysRevLett.109.103602}%
  \BibitemOpen
  \bibfield  {author} {\bibinfo {author} {\bibfnamefont {Shamir}\ \bibnamefont {Rosen}}, \bibinfo {author} {\bibfnamefont {Itai}\ \bibnamefont {Afek}}, \bibinfo {author} {\bibfnamefont {Yonatan}\ \bibnamefont {Israel}}, \bibinfo {author} {\bibfnamefont {Oron}\ \bibnamefont {Ambar}}, \ and\ \bibinfo {author} {\bibfnamefont {Yaron}\ \bibnamefont {Silberberg}},\ }\bibfield  {title} {\enquote {\bibinfo {title} {Sub-rayleigh lithography using high flux loss-resistant entangled states of light},}\ }\href {\doibase 10.1103/PhysRevLett.109.103602} {\bibfield  {journal} {\bibinfo  {journal} {Phys. Rev. Lett.}\ }\textbf {\bibinfo {volume} {109}},\ \bibinfo {pages} {103602} (\bibinfo {year} {2012})}\BibitemShut {NoStop}%
\bibitem [{\citenamefont {Ladd}\ \emph {et~al.}(2006)\citenamefont {Ladd}, \citenamefont {van Loock}, \citenamefont {Nemoto}, \citenamefont {Munro},\ and\ \citenamefont {Yamamoto}}]{Ladd2006}%
  \BibitemOpen
  \bibfield  {author} {\bibinfo {author} {\bibfnamefont {T.~D.}\ \bibnamefont {Ladd}}, \bibinfo {author} {\bibfnamefont {P.}~\bibnamefont {van Loock}}, \bibinfo {author} {\bibfnamefont {K.}~\bibnamefont {Nemoto}}, \bibinfo {author} {\bibfnamefont {W.~J.}\ \bibnamefont {Munro}}, \ and\ \bibinfo {author} {\bibfnamefont {Y.}~\bibnamefont {Yamamoto}},\ }\bibfield  {title} {\enquote {\bibinfo {title} {Hybrid quantum repeater based on dispersive {CQED} interactions between matter qubits and bright coherent light},}\ }\href {\doibase 10.1088/1367-2630/8/10/184} {\bibfield  {journal} {\bibinfo  {journal} {New Journal of Physics}\ }\textbf {\bibinfo {volume} {8}},\ \bibinfo {pages} {184} (\bibinfo {year} {2006})}\BibitemShut {NoStop}%
\bibitem [{\citenamefont {Israel}\ \emph {et~al.}(2014)\citenamefont {Israel}, \citenamefont {Rosen},\ and\ \citenamefont {Silberberg}}]{PhysRevLett.112.103604}%
  \BibitemOpen
  \bibfield  {author} {\bibinfo {author} {\bibfnamefont {Yonatan}\ \bibnamefont {Israel}}, \bibinfo {author} {\bibfnamefont {Shamir}\ \bibnamefont {Rosen}}, \ and\ \bibinfo {author} {\bibfnamefont {Yaron}\ \bibnamefont {Silberberg}},\ }\bibfield  {title} {\enquote {\bibinfo {title} {Supersensitive polarization microscopy using {NOON} states of light},}\ }\href {\doibase 10.1103/PhysRevLett.112.103604} {\bibfield  {journal} {\bibinfo  {journal} {Phys. Rev. Lett.}\ }\textbf {\bibinfo {volume} {112}},\ \bibinfo {pages} {103604} (\bibinfo {year} {2014})}\BibitemShut {NoStop}%
\bibitem [{\citenamefont {Meher}\ \emph {et~al.}(2024)\citenamefont {Meher}, \citenamefont {Poem}, \citenamefont {Opatrn\'y}, \citenamefont {Firstenberg},\ and\ \citenamefont {Kurizki}}]{PhysRevA.110.013715}%
  \BibitemOpen
  \bibfield  {author} {\bibinfo {author} {\bibfnamefont {Nilakantha}\ \bibnamefont {Meher}}, \bibinfo {author} {\bibfnamefont {Eilon}\ \bibnamefont {Poem}}, \bibinfo {author} {\bibfnamefont {Tom\'a\ifmmode \check{s}\else~\v{s}\fi{}}\ \bibnamefont {Opatrn\'y}}, \bibinfo {author} {\bibfnamefont {Ofer}\ \bibnamefont {Firstenberg}}, \ and\ \bibinfo {author} {\bibfnamefont {Gershon}\ \bibnamefont {Kurizki}},\ }\bibfield  {title} {\enquote {\bibinfo {title} {Supersensitive phase estimation by thermal light in a {K}err-nonlinear interferometric setup},}\ }\href {\doibase 10.1103/PhysRevA.110.013715} {\bibfield  {journal} {\bibinfo  {journal} {Phys. Rev. A}\ }\textbf {\bibinfo {volume} {110}},\ \bibinfo {pages} {013715} (\bibinfo {year} {2024})}\BibitemShut {NoStop}%
\bibitem [{\citenamefont {Vogel}\ and\ \citenamefont {Sperling}(2014)}]{PhysRevA.89.052302}%
  \BibitemOpen
  \bibfield  {author} {\bibinfo {author} {\bibfnamefont {W.}~\bibnamefont {Vogel}}\ and\ \bibinfo {author} {\bibfnamefont {J.}~\bibnamefont {Sperling}},\ }\bibfield  {title} {\enquote {\bibinfo {title} {Unified quantification of nonclassicality and entanglement},}\ }\href {\doibase 10.1103/PhysRevA.89.052302} {\bibfield  {journal} {\bibinfo  {journal} {Phys. Rev. A}\ }\textbf {\bibinfo {volume} {89}},\ \bibinfo {pages} {052302} (\bibinfo {year} {2014})}\BibitemShut {NoStop}%
\end{thebibliography}
\end{document}